\documentclass[final,authoryear,3p,times]{elsarticle}

\usepackage{subcaption}
\usepackage[percent]{overpic}
\usepackage{graphicx,color}
\usepackage{enumerate}
\usepackage[colorlinks]{hyperref}
\usepackage{amsmath,amssymb,amsthm,bm}
\usepackage{dsfont}
\usepackage{float}
\usepackage{multicol}
\usepackage[dvipsnames]{xcolor}
\usepackage{ulem}
\usepackage{tikz}
\usetikzlibrary{automata,positioning}
\usepackage{afterpage}
\usepackage{calc}
\usepackage{ifthen}
\usepackage{algorithm}
\usepackage{algpseudocode}
\usepackage{caption}
\usepackage{subcaption}
\usepackage{array}
\usepackage{cleveref}

\newcommand{\boldface}[1]{\boldsymbol{#1}}  %

\newcommand{\bfe}{\boldface{e}}

\newcommand{\bfh}{\boldface{h}}

\newcommand{\bfy}{\boldface{y}}

\newcommand{\bfH}{\boldface{H}}
\newcommand{\bfI}{\boldface{I}}

\newcommand{\bfX}{\boldface{X}}

\newcommand{\bfepsvar}{\boldsymbol{\epsilon}}

\newcommand{\bfgamma}{\boldsymbol{\gamma}}
\newcommand{\bfdelta}{\boldsymbol{\delta}}

\newcommand{\bfmu}{\boldsymbol{\mu}}

\newcommand{\bfXi}{\boldsymbol{\Xi}}

\newcommand{\calD}{\mathcal{D}}

\newcommand{\calF}{\mathcal{F}}
\newcommand{\calG}{\mathcal{G}}
\newcommand{\calH}{\mathcal{H}}

\newcommand{\calN}{\mathcal{N}}

\newcommand{\calU}{\mathcal{U}}

\newcommand{\Rset}{\mathbb{R}}

\newlength{\boxwidth}
\def\dd{\;\!\mathrm{d}}

\def\btheorem{\begin{theorem}}
\def\etheorem{\end{theorem}}
\def\blemma{\begin{lemma}}
\def\elemma{\end{lemma}}
\def\bproposition{\begin{proposition}}
\def\eproposition{\end{proposition}}
\def\bcorollary{\begin{corollary}}
\def\ecorollary{\end{corollary}}
\def\bdefinition{\begin{definition}}
\def\edefinition{\end{definition}}
\def\bexample{\begin{example}}
\def\eexample{\end{example}}
\def\bremark{\begin{remark}}
\def\eremark{\end{remark}}

\newcommand{\be}{\begin{equation}}
\newcommand{\ee}{\end{equation}}
\newcommand{\beq}{\begin{eqnarray}}
\newcommand{\eeq}{\end{eqnarray}}
\newcommand{\bem}{\begin{multline}}
\newcommand{\eem}{\end{multline}}
\newcommand{\ba}{\begin{align}}
\newcommand{\ea}{\end{align}}

\journal{arXiv}

\begin{document}

\begin{frontmatter}

\title{Data-efficient continuous conditional denoising diffusion model for microstructure generation}

\author[a]{Tarakram Ramgopal}
\author[a]{Gowtham Nimmal Haribabu}
\author[a,b]{Hussein Farahani}
\author[a,b]{Cornelis Bos}
\author[a]{Siddhant Kumar}

\ead{Sid.Kumar@tudelft.nl}
\address[a]{Department of Materials Science and Engineering, Delft University of Technology, 2628 CD Delft, The Netherlands}
\address[b]{Tata Steel, Research \& Development, P.O. Box 10000, IJmuiden, 1970 CA, Netherlands}

\begin{abstract}
Traditional computational models, such as cellular automata and phase-field methods, are effective for simulating microstructural evolution but often face computational bottlenecks, limiting their application in high-throughput and on-demand process optimization. Generative machine learning approaches, such as denoising diffusion models, have emerged as powerful tools for surrogate modeling of process-structure maps, specifically producing representative microstructures conditioned on process parameters. However, they often require large amounts of data for training, particularly when process conditions are continuous rather than discrete categorical variables. To address this, we present a continuous conditional denoising diffusion model for generating microstructures conditioned on processing parameters. Trained on a compact dataset of process-microstructure pairs, this framework first adds noise to microstructure images and then trains a neural network to progressively remove the noise, learning the underlying statistical patterns of the microstructure. To address data inefficiencies associated with continuously valued process conditions, we propose a vicinal-loss training strategy that associates process conditions in data-sparse regions with nearby conditions in the dataset. Combined with classifier-free guidance and denoising diffusion implicit sampling, this approach enables data-efficient continuous conditional generation of microstructures compared to classical denoising diffusion models. The model successfully generates representative microstructures for low-carbon steel conditioned on manganese composition, matching key physical features such as phase and grain morphology, grain size distribution, phase fraction, and interfacial area distribution. More generally, this approach opens avenues for efficient process design and optimization of materials and their microstructures.
\end{abstract}

\end{frontmatter}

\section{Introduction}
\label{sec:Introduction}

In metallic systems, processing parameters influence microstructural evolution, which in turn dictates material properties and performance. Conversely, achieving tailored properties requires control of the microstructure, which in turn necessitates process optimization. However, traditional trial-and-error methods for constructing process-structure maps are inefficient due to high experimental costs. While simulations based on, e.g., cellular automata \citep{Yazdipour2008,Bos2010,Bos2011},  phase field methods \citep{Chen2002,Choi2024, Bhadeshia2014,Peivaste2022, Hu2022, Gao2023, Xue2022, MontesdeOcaZapiain2021}, and  Potts-type Monte Carlo simulations \citep{Hore2013, Tong2002} offer an efficient means to virtually explore process–structure relationships, their long computational runtimes often preclude real-time, process-conditioned microstructure generation and process optimization required in industrial metals production and processing. For example, a key circularity challenge is scrap steel recycling, where varying impurity levels affect material composition, requiring models that relate processing parameters to microstructural features for real-time control to achieve target properties. In such cases, long simulation times are impractical, and high-throughput, process-conditioned microstructure generation is required.

\textbf{Deep learning for microstructure generation:} In recent years, machine learning (ML), specifically deep neural networks, has emerged as a promising route for accelerated microstructure modeling. Models based on recurrent networks \citep{Hu2022, MontesdeOcaZapiain2021}, which capture the history dependence of microstructure evolution, and convolutional networks \citep{Gao2023, Choi2024}, which encode spatial structure in image-based microstructure representations, have demonstrated considerable success. Following an offline training stage, fast GPU-compatible implementations of neural networks can yield significant computational speedups relative to physics-based models during the inference stage \citep{Mordvintsev2020, Tang2023, Seibert2024}. More recently, graph networks \citep{Xue2022, Qin2023} have also shown promise in improving data and computational efficiency by replacing image-based representations with grain-level representations encoded as graphs. 

\textbf{Discriminative models are incompatible with stochastic microstructures:} Earlier ML models for microstructure generation \citep{Peivaste2022, MontesdeOcaZapiain2021, Tang2023}  were predominantly discriminative in nature, in the sense that, given an initial microstructure, the model learns to predict the microstructure at a future time instance by minimizing a loss between the ground truth and the predicted microstructure. This discriminative approach inherently assumes the existence of a unique true microstructure at the future time instance. However, this assumption does not hold in the presence of stochastic processes, such as nucleation, which introduce new grains at intermediate times. Since nucleation is inherently stochastic, both spatially and temporally, it invalidates the discriminative approach, rendering such models unsuitable for microstructure generation scenarios in which nucleation occurs.

\textbf{Generative modeling for stochastic microstructure generation:} To address the limitations of discriminative approaches, recent fundamental ML methods \citep{goodfellow2014gan, Kingma2013, Dinh2014,  Rezende2015, Nichol2021,Kingma2021} take a generative approach, whereby the ML model learns the probabilistic distribution of the data. In the context of process-structure relationships and microstructure generation (see e.g., \cite{Azqadan2023}), this corresponds to explicitly or implicitly learning the conditional probability, i.e., the probability distribution of the microstructures conditioned on a given set of process conditions, and enabling sampling from this distribution. The generated microstructures need not exactly match those in the dataset but should exhibit features that are statistically representative of it. The generative model is trained by maximizing the conditional likelihood of generating microstructures from the dataset, in contrast to the discriminative approach, and can therefore accommodate generation of stochastic microstructures.

A generative model for microstructures should exhibit three key attributes: \textit{(i)} high quality or fidelity of the generated microstructures, \textit{(ii)} efficient sampling speed, and \textit{(iii)} strong mode coverage or diversity in the generated microstructures. At the time of writing, amid a rapidly progressing field, there are three main categories of generative models (both in the broader ML context and in their application to microstructure generation); each model typically performs well in only two of these three aspects.
\begin{itemize}
    \item \textbf{Generative adversarial networks} (GANs) \citep{goodfellow2014gan} and their application in microstructure modeling \citep{Hsu2021, Chun2020, Tang2021, Haribabu2023} can produce high-quality, high-fidelity microstructures with fast sampling speeds, but suffer from mode collapse, i.e., limited diversity in the generated microstructures.
    \item \textbf{Variational autoencoders} (VAEs) and \textbf{normalizing flows} (NFs) \citep{Kingma2013, Rezende2015, Dinh2014} for microstructure modeling \citep{Kim2021, White2024, Ji2024, Mirzaee2025} provide improved mode coverage (i.e., higher diversity) with fast sampling speeds, but at the cost of reduced quality and fidelity compared to GANs.
    \item \textbf{Denoising diffusion models} \citep{Kingma2021, Ho2020,Nichol2021} are generally considered the state-of-the-art in microstructure modeling \citep{Vlassis2024, FernandezZelaia2024, Yang2026,  Lyu2024, Buzzy2024, Azqadan2023, Dureth2023, Bentamou2025, Baishnab2025}, offering high quality and fidelity along with extensive mode coverage and diversity, but they suffer from slow sampling speeds compared to GANs, VAEs, and NFs. Recent methods, such as denoising diffusion implicit modeling (DDIM) \citep{Song2020} and GAN-based distillation models \citep{Yin2023,Yin2024} trained against pre-trained diffusion models, are promising directions for accelerating sampling in diffusion models.
\end{itemize}
In this work, we focus on diffusion models for the conditional generation of microstructures. Although diffusion models have demonstrated success in microstructure generation, we identify three challenges and propose a framework of solutions to address them.
\begin{enumerate}
    
    \item \textbf{Data efficiency for continuous conditioning:} Diffusion models are typically built upon deep and wide neural network architectures with a large number of trainable parameters and are therefore data-hungry during training. This problem is further exacerbated in conditional microstructure generation compared with unconditional generation, as learning the conditional probability distribution requires more data than learning the unconditional probability of the dataset. Furthermore, within conditional microstructure generation, prior works have primarily focused on discrete conditionals, i.e., process conditions that are discrete categorical variables \citep{Azqadan2023, Dureth2023}. In contrast, continuous conditioning, where the process conditions can take continuous real-valued values, remains relatively underexplored, and the associated data requirements become extremely high.  
    This raises the question -- \textit{how to improve the data efficiency of diffusion models for continuous conditioning in microstructure generation?} To address this, we propose a training strategy based on vicinal loss \citep{Ding2020,Ding2024} that improves data efficiency for training with continuous conditioning.
    
    \item \textbf{Sampling speed:} As mentioned earlier, classical diffusion models suffer from slow sampling speeds compared to GANs, VAEs, and NFs. Achieving fast sampling speed is an important step toward real-time process optimization in real-world applications. To address this issue, we explore the viability of DDIM for accelerated conditional microstructure generation.
    
    \item \textbf{Quality of conditional generation:} While classical diffusion models offer good quality and diversity in generated microstructures, the quality of the generated samples can degrade under continuous conditioning, especially when the available data are limited. This issue is particularly relevant for microstructures, as opposed to general image generation, because localized features such as grain boundaries -- typically only a few pixels wide but forming closed loops spanning tens to thousands of pixels -- are difficult to capture accurately. To address this challenge, we explore classifier-free guidance (CFG) \citep{Ho2022} as a technique to improve conditional generation toward higher-quality microstructures.
\end{enumerate}

Our framework, comprising vicinal loss, DDIM, and CFG, achieves higher quality conditional microstructure generation at faster sampling speeds compared to classical diffusion models. We demonstrate this through a series of ablation studies in which each component is selectively disabled to study its contribution. As a prototypical system, we consider low-carbon steel microstructures conditioned on manganese concentration, with data generated from cellular automata simulations. We note that the proposed framework is data-agnostic, i.e., it can be generalized to other microstructures and, more broadly, to metallic systems, as well as to data generated using other, more computationally expensive, physics-based solvers such as phase-field or Monte Carlo simulations.

This paper is organized as follows. Section \ref{sec:ccddm_mc_gen} introduces the formulation of the continuous conditional diffusion model. Specifically,  Section \ref{subsec:problem_setting} presents the problem setting, followed by details of data generation in Section \ref{subsec:data_generation}. Section \ref{subsec:diffusion_models} presents the mathematical formulation of a classical diffusion model. Section \ref{subsec:cond_emb} and Section \ref{subsec:vicinal-loss} introduce strategies for continuous process-condition embedding and data-efficient training using vicinal loss. Section \ref{subsec:model_arch} describes the neural network model architecture. Section \ref{subsec:cfg_ddim} presents strategies for high-quality and efficient sampling using CFG and DDIM. Section \ref{sec:results} summarizes the performance of the framework for continuous conditional microstructure generation through qualitative and quantitative analyses of the generated microstructures and includes a series of ablation studies to assess the role of each proposed enhancement strategy. Finally, Section \ref{sec:Conclusion} concludes the paper.

\section{Continuous conditional diffusion model for microstructure generation}
\label{sec:ccddm_mc_gen}

\subsection{Problem setting} 
\label{subsec:problem_setting}

We consider low-carbon steels that start as fully austenitic at high temperature and, upon cooling, transform to microstructures dominated by a ferrite matrix along with a secondary phase -- each represented by an image of dimensions $C\times H\times W$, where $C$ is the number of channels, $H$ is the image height, and $W$ is the image width. For a comprehensive description of multi-phase microstructures, the number of channels corresponds to the number of phases and can be greater than one. In the context of this work, we focus on two phases (austenite and ferrite) and introduce a compact single-channel image representation, while noting that the generative modeling method is generalizable to multi-channel microstructure representations. 

\begin{figure}
    \centering
    \includegraphics[width=0.95\textwidth]{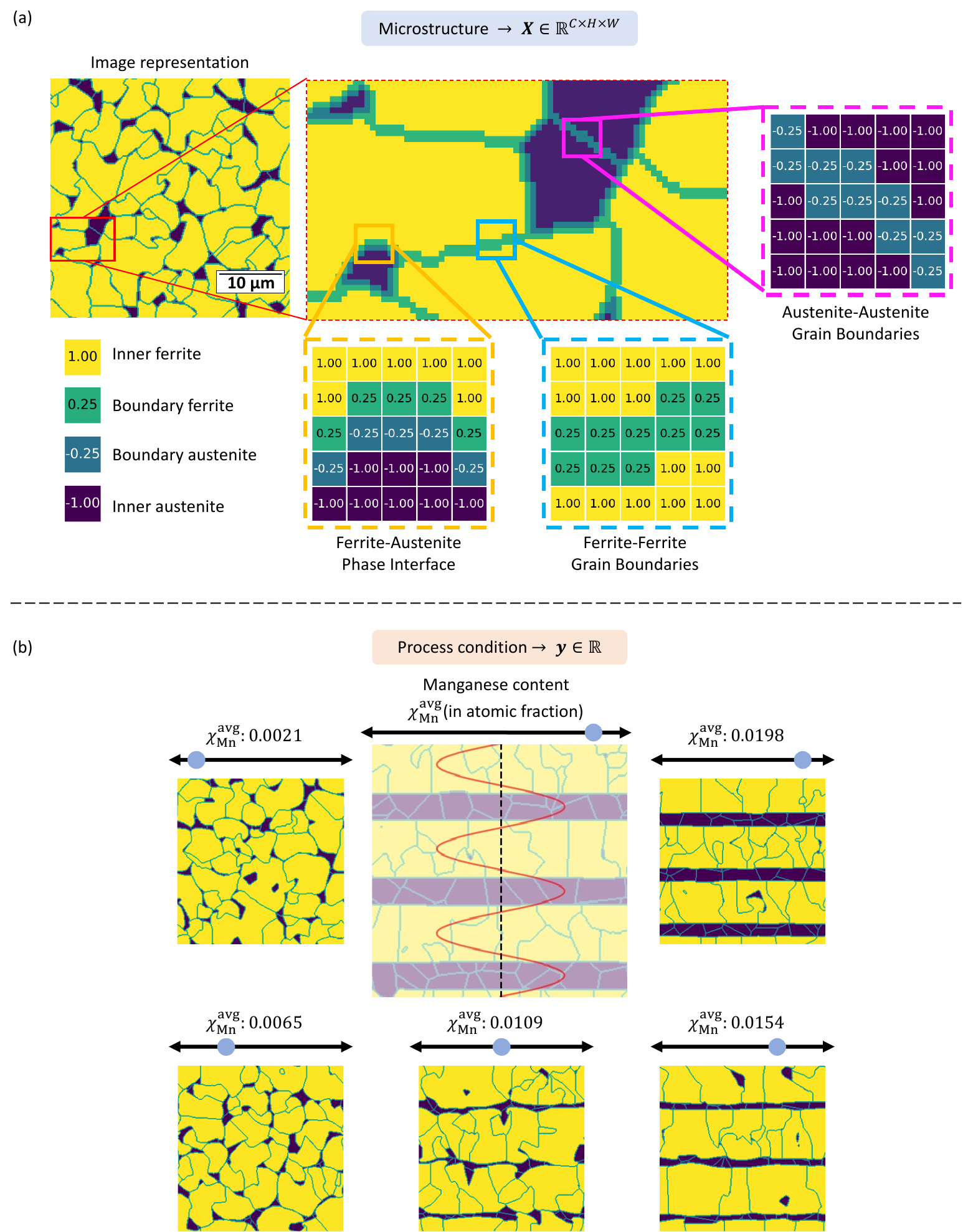}
    \caption{(a) Schematic of the image representation used for two-dimensional low-carbon steel microstructures, with insets showcasing phase interfaces and grain boundaries. (b) Schematic of the manganese concentration distribution  (see \eqref{eq:Mn_effect}) controlled by the process condition $\chi^\text{avg}_\text{Mn}$. The arrow with blue dot indicates the value of $\chi^\text{avg}_\text{Mn}$ within the range defined by the minimum and maximum average manganese concentrations in the dataset. Representative examples of steel microstructures corresponding to different average manganese concentrations are also shown.
    }
    \label{fig:dataset_description}
\end{figure}

Specifically, we consider an image $\bfX\in\{-1,-0.25,0.25,1\}^{C\times H\times W}$ with $C=1$ to represent a microstructure.  Cells with values $-1$ and $1$ correspond to the austenite and ferrite phases, respectively. In order to distinguish boundaries, we assign cell values of $-0.25$ and $0.25$ to austenite and ferrite boundaries, respectively. A ferrite-austenite interface is defined by a pair of neighboring cells with values of $-0.25$ and $0.25$. An austenite-austenite grain boundary consists of two neighboring cells, each with a value of $-0.25$. Similarly, a ferrite-ferrite grain boundary is characterized by two neighboring cells, each having a value of $0.25$. Figure \ref{fig:dataset_description} illustrates this channel description. 

This convention allows encoding all the information needed to extract ferrite and austenite grains (via thresholding and image segmentation algorithms during post-processing), along with precise grain boundaries, using a single channel---eliminating the need for multiple channels. The numerical values are distinct enough to be reliably recognized by a ML model. Additionally, using a single channel enables efficient microstructure representation for generative ML, where computational cost increases significantly with the number of channels.

To decode process-structure relationships, we consider a sufficiently large and representative dataset of process-structure pairs
\be
\calD = \left\{ \left( \bfX_i, \bfy_i \right): \; i = 1, 2, \dots, N \right\},
\ee
where $\bfX_i$ represents a microstructure corresponding to process conditions $\bfy_i\in\Rset^P$ (with dimensionality $P$). Here, $N$ is the number of data points and the subscript $(\cdot)_i$ refers to the $i^\text{th}$ data point in $\calD$. A generative ML model trained on this dataset aims to generate a microstructure image $\hat\bfX = \calG(\bfy)$ as a function of an arbitrary process condition $\bfy$. Process conditions---such as alloy composition or cooling rates---serve as inputs, guiding the ML model to generate microstructural images that accurately reflect the corresponding morphological and statistical features.  Here, we consider manganese concentration as a representative process parameter to control the austenite to ferrite phase transformation; other process parameters may include carbon concentration or cooling schedule.

\subsection{Data generation}
\label{subsec:data_generation}

We use Cellular Automaton Sharp Interface Phase Transformation (CASIPT) \citep{Bos2010, Bos2011} as the basis for microstructure evolution simulations and dataset generation. We simulate the transformation of austenite to ferrite based on a mixed-mode model \citep{Bos2007} on a two-dimensional grid with cell size $\Delta d$ and a dynamic time discretization approach that maintains a sufficiently small timestep to obey physical constraints and limit the austenite/ferrite interface movement to no more than one grid cell per iteration. Each cell embeds a set of properties such as phase (austenite, ferrite), unique grain IDs, carbon concentration, crystallographic orientation, and other relevant information to determine cell response during carbon diffusion, grain growth, and phase transformation. 

We start from an initial austenite microstructure that features a homogeneous distribution of carbon and silicon at average concentrations $\chi^\text{avg}_\text{C}$ and $\chi^\text{avg}_\text{Si}$, respectively, throughout the grid. We simulate austenite to ferrite cooling starting at an initial temperature $T_\text{initial}$ then cooling at rate $\phi_1$ for time $\Delta{t}_1$, followed by cooling at rate $\phi_2$ for time $\Delta{t}_2$, and finally an isothermal holding for time $\Delta{t}_3$. We assume a spatially varying manganese banding profile (see Figure \ref{fig:dataset_description} for a schematic illustration) typical of steel production processes. The manganese concentration ($\chi_\text{Mn}$) varies sinusoidally along the $Y$-direction while remaining constant along the $X$-direction. At any given point, we define it as
\be\label{eq:Mn_effect}
    \chi_{\text{Mn}}(y_\text{domain}) = \chi_\text{Mn}^\text{avg} + A \sin\left(\frac{y_{\text{domain}}}{F}\right)
    \qquad \text{with} \quad F = \frac{n_y \Delta d}{2N_\text{periods}},
\ee
where {$\chi^\text{avg}_\text{Mn}$ is the average manganese concentration in the range of $0.0021-0.0220$ atomic fraction} and represents the process condition $\bfy_i\in\Rset$ for the ML model. $y_\text{domain}$ is the pixel location along $Y$-direction, $A$ is the amplitude of the sinusoidal variation, $F$ is the length-scale parameter of the sine wave with  $N_\text{periods}$ sinusoidal periods along the $Y$-direction, $n_y$ is the total length of the system in the $Y$-direction, and $\Delta d$ is the cell size. \ref{sec:parameters} summarizes and provides the parameter values used for data generation. By varying $\chi^\text{avg}_\text{Mn}$, we vary the austenite banding effect in the microstructure; representative examples are shown in Figure \ref{fig:dataset_description}. Note that all parameters, except $\chi^\text{avg}_\text{Mn}$, remain constant in this study.

The dataset $\calD$ is generated by randomly sampling $\chi^\text{avg}_\text{Mn}$ (see \ref{sec:Xmn_distribution} for sampling strategy) and obtaining the corresponding microstructure via CASIPT. Note that we randomly generate the initial austenite microstructure with isotropic texture. Consequently, the resulting austenite–ferrite microstructures may differ even under identical process parameters, although they exhibit statistically similar morphological characteristics.  

In the subsequent section, we present the formulation of the ML model $\calG$.

\subsection{Data-efficient denoising diffusion model}
\label{subsec:diffusion_models}

\begin{figure}
\centering
\includegraphics[width=\textwidth]{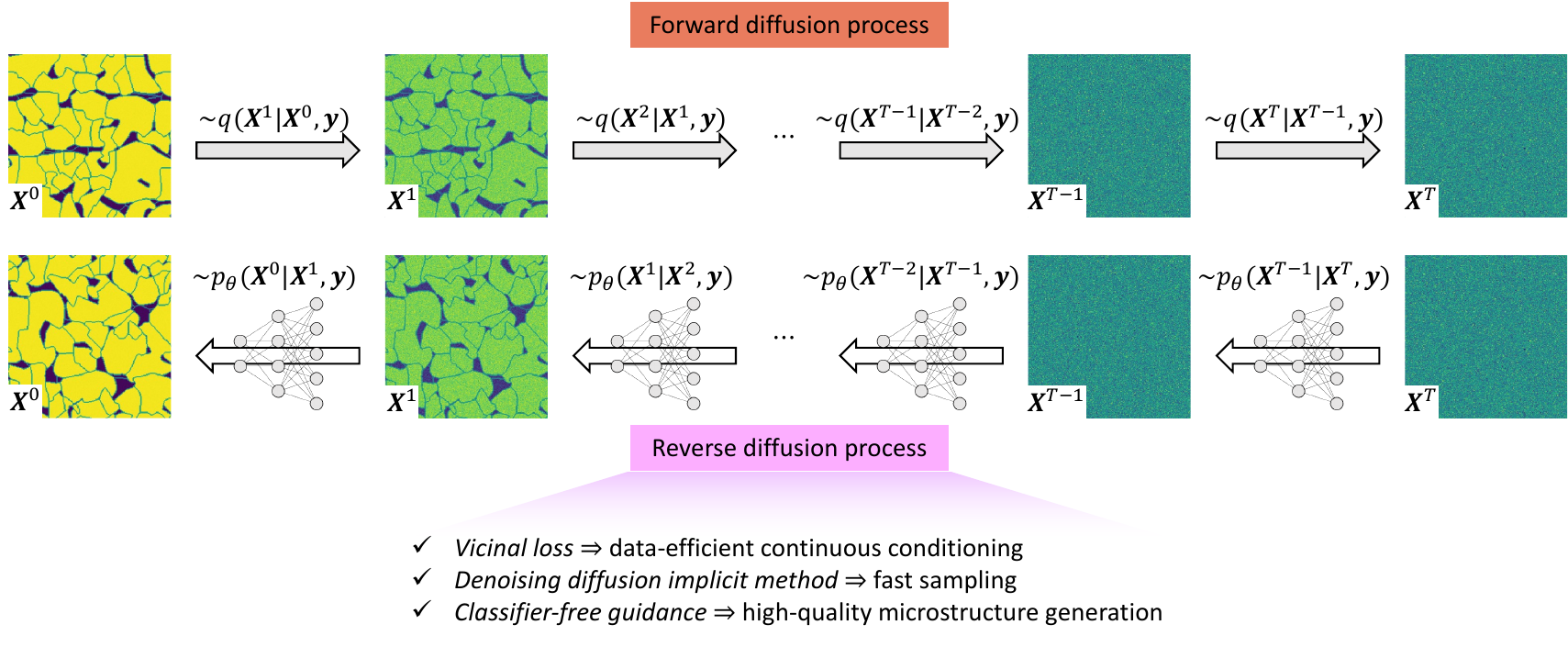}
    \caption{Schematic of the denoising diffusion model, where the forward diffusion process is performed using $q(\bfX^t \mid \bfX^{t-1}, \bfy)$ for $t\in\{1,\dots,T\}$, and the reverse diffusion process is parameterized using a NN: $p_\theta(\bfX^{t-1} \mid \bfX^t, \bfy)$ with the set of trainable weights $\theta$.}\label{fig:overview}
\end{figure}

We adapt the denoising diffusion model \citep{Ho2020, Luo2022} for image generation in computer vision setting to microstructures context. The core concept of a diffusion model is to progressively add noise to dataset images through a \textbf{forward diffusion process} until all information content is lost. A neural network (NN) is then trained to perform the \textbf{reverse diffusion process}, learning to progressively denoise and reconstruct the images step by step. Figure \ref{fig:overview} provides a schematic of the model framework.

\textbf{Forward diffusion process}: Let $\bfX^0$ be a microstructural image associated with process condition $\bfy$ from the dataset $\calD$. The forward diffusion process is a Markov chain described by the probability distribution $q$ given by
\be\label{eq:fwd_process}
q(\bfX^t \mid \bfX^{t-1}, \bfy) = \calN(\bfX^t; \sqrt{1-\beta_t} \, \bfX^{t-1}, \ \beta_t \bfH_y).
\ee
Here,  $\bfX^t$ denotes the noisy microstructure image at the $t^\text{th}$ step of the Markov chain with $t\in\{1,\dots,T\}$.\footnote{Note that this iterative stepping is only in the context of the Markov chain of the diffusion model and does not concern the physical temporal evolution of the microstructures.} $\calN(\bfX; \ \bm{\mu}, \ \bm{\Xi})$ denotes the multivariate Gaussian distribution in $\bfX$ with mean $\bfmu$ and covariance $\bm{\Xi}$. The covariance is designed to communicate the influence of process condition $\bfy$ in form of covariance matrix $\bfH_y$ during the forward (and reverse) diffusion process (see Section \ref{subsec:cond_emb} for motivation and implementation details on $\bfH_y$). Additionally, $0\leq\beta_1<\beta_2<\dots<\beta_t<\dots\beta_T\leq 1$ are scalar hyperparameters  that determine the amount of variance (i.e., noising) added at each step of the forward diffusion (see \ref{sec:noise_scheduler} for implementation details on noise scheduler). The number of steps $T$ is chosen to be sufficiently large so that $\bfX^T$ becomes highly noisy with no meaningful information content, closely resembling $\bfy$-dependent Gaussian noise $\calN(\bm{0},\bfH_y)$.

Leveraging the Markov property, the probability of the noisy image at any step $t$ given $\bfX^0$ is obtained as
\be\label{eqn:x^t}
q(\bfX^t\mid\bfX^0,\bfy) =  \prod_{\tau=1}^t q(\bfX^\tau\mid\bfX^{\tau-1},\bfy) = \calN(\bfX^t; \sqrt{\bar{\alpha}_t} \bfX^0 , (1 - \bar{\alpha}_t)\bfH_y), \qquad \text{with}\quad \bar{\alpha}_t = \prod_{i=1}^t \alpha_i,\quad \text{and}\quad   \alpha_t = 1 - \beta_t.
\ee
The above can be reparameterized as
\be\label{eq:forward-reparam}
\bfX^t = \sqrt{\bar{\alpha}_t} \bfX^0 + \sqrt{1 - \bar{\alpha}_t} \bfepsvar, \qquad \text{with}\quad \bfepsvar \sim \calN(\bm{0}, \bfH_y).
\ee

\textbf{Reverse diffusion process}:  We aim to acquire $\bfX^0(\bfy)$ from a {$\bfy$-dependent} Gaussian noise $\bfX^T\sim\calN(\bm{0},\bfH_y)$ for given process condition~$\bfy$. However, estimating the inverse of the forward process directly, i.e., $q(\bfX^{t-1} \mid \bfX^t,\bfy)$ and consequently, the final microstructure image via $q(\bfX^0 \mid \bfX^T,\bfy)$ are computationally intractable. 

To this end, using known $q(\bfX^t \mid \bfX^0,\bfy)$ and $q(\bfX^{t-1} \mid \bfX^0,\bfy)$ as well as Bayes' rule, we can obtain the conditional posterior probability for $t\geq2 $ (see \cite{Luo2022} for detailed derivations)
\be\label{eq:q-reverse}
q(\bfX^{t-1} \mid \bfX^t, \bfX^0, \bfy) = \calN(\bfX^{t-1};\bfmu_q(\bfX^t,\bfX^0,\bfy),\bfXi_q(t))
\ee
with
\be\label{eq:q-reverse-mean}
\bfmu_q(\bfX^t,\bfX^0,\bfy) =  \frac{\sqrt{\alpha_t} (1 - \bar{\alpha}_{t-1})}{1 - \bar{\alpha}_t} \bfX^t + \frac{\sqrt{\bar{\alpha}_{t-1}}(1-\alpha_t)}{1 - \bar{\alpha}_t} \bfX^0
\ee
and
\be\label{eq:q-reverse-variance}
\bfXi_q(t) = \frac{(1-\alpha_t)(1-\bar\alpha_{t-1})}{1-\bar\alpha_t}\bfH_y.
\ee
The above probability is conditioned upon $\bfX^0$ which is inherently unknown. 

We introduce a reverse diffusion process parameterized by a NN -- with the set of trainable weights $\theta$ -- to denoise a noisy image $\bfX^t$ from step $t$ to $(t-1)$ following a probability distribution $p_\theta$ of the form:
\be\label{eq:ansatz}
p_\theta(\bfX^{t-1} \mid \bfX^t, \bfy) = \calN(\bfX^{t-1};\bfmu_\theta(\bfX^t, t, \bfy), \bfXi_q(t)),
\ee
with
\be\label{eq:ansatz_mean}
\bfmu_\theta(\bfX^t,t,\bfy) =  \frac{\sqrt{\alpha_t} (1 - \bar{\alpha}_{t-1})}{1 - \bar{\alpha}_t} \bfX^t + \frac{\sqrt{\bar{\alpha}_{t-1}}(1-\alpha_t)}{1 - \bar{\alpha}_t} \calF_\theta(\bfX^t,t,\bfy).
\ee
In the ansatz \eqref{eq:ansatz} for the reverse diffusion, we assume the same form for the mean and covariance as in the reverse of the forward diffusion \eqref{eq:q-reverse} (see \eqref{eq:q-reverse-mean} and \eqref{eq:q-reverse-variance}, respectively). $\calF_\theta(\bfX^t,t,\bfy)$ is a denoising NN that takes as input the noisy image $\bfX^t$ (and corresponding step $t$ and process condition $\bfy$) and outputs a clean image signal. The denoising in \eqref{eq:ansatz} is carried out iteratively from step $t=T$ to $t=1$ to obtain the final microstructure at $t=0$, i.e., 
\be
\hat\bfX = \calG(\bfy) = \bfX^0(\bfy),\qquad \text{with} \quad \bfX^{t-1}\sim p_\theta(\bfX^{t-1}\mid\bfX^t,\bfy) \quad \forall \quad t=1,\dots,T.
\ee

In the subsequent sections, we elaborate on different embedding schemes, training protocol, architecture, and sampling strategy for the NN $\calF_\theta (\bfX^t,t,\bfy)$. 

\subsection{Process condition embedding}
\label{subsec:cond_emb}

The process condition $\bfy$ (the average manganese concentration $\chi^\text{avg}_\text{Mn} \in \Rset$ in this case) is continuous and essentially in an infinite space (as opposed to traditional class  or categorical conditioning). In light of data scarcity, we need to obtain efficient embedding schemes for $\bfy$ before training a continuous conditional diffusion model.

The process parameters $\bfy$ influence the diffusion model through two routes: {conditioning in the covariance $\bfH_y$ and conditioning in the denoising NN $\calF_\theta(\bfX^t,t,\bfy)$}. This is achieved by creating high-dimensional embeddings of $\bfy$ for both routes.

We model $\bfH_y$ as a diagonal covariance matrix, i.e., 
\be
\bfH_y = \text{diag}(\text{exp}(-\bfh^{\bm{\Xi}}))\in\Rset^{d^{\bm{\Xi}}\times d^{\bm{\Xi}}}
\ee
and sample $\bfy$-dependent Gaussian noise $\bfepsvar\sim\calN(\bm{0}, \bfH_y)$ to modulate the forward diffusion process \eqref{eq:fwd_process} and reverse diffusion process \eqref{eq:ansatz}. Here, $\bfH_y$ is reshaped from a high-dimensional \textit{covariance condition embedding} $\bfh^{\bm{\Xi}}$ with dimensionality $d^{\bm{\Xi}}$ as the length of the flattened microstructure image $\bfX$ in the dataset (i.e. $d^{\bm{\Xi}} = (C\times H\times W)$).

We create \textit{continuous condition embedding} $\bfh\in \Rset^d$ with dimensionality $d$ to effectively pass process condition $\bfy$ as an input to the model $\calF_\theta(\bfX^t,t,\bfy)$.

\begin{figure}
    \centering \includegraphics[width=0.95\textwidth]{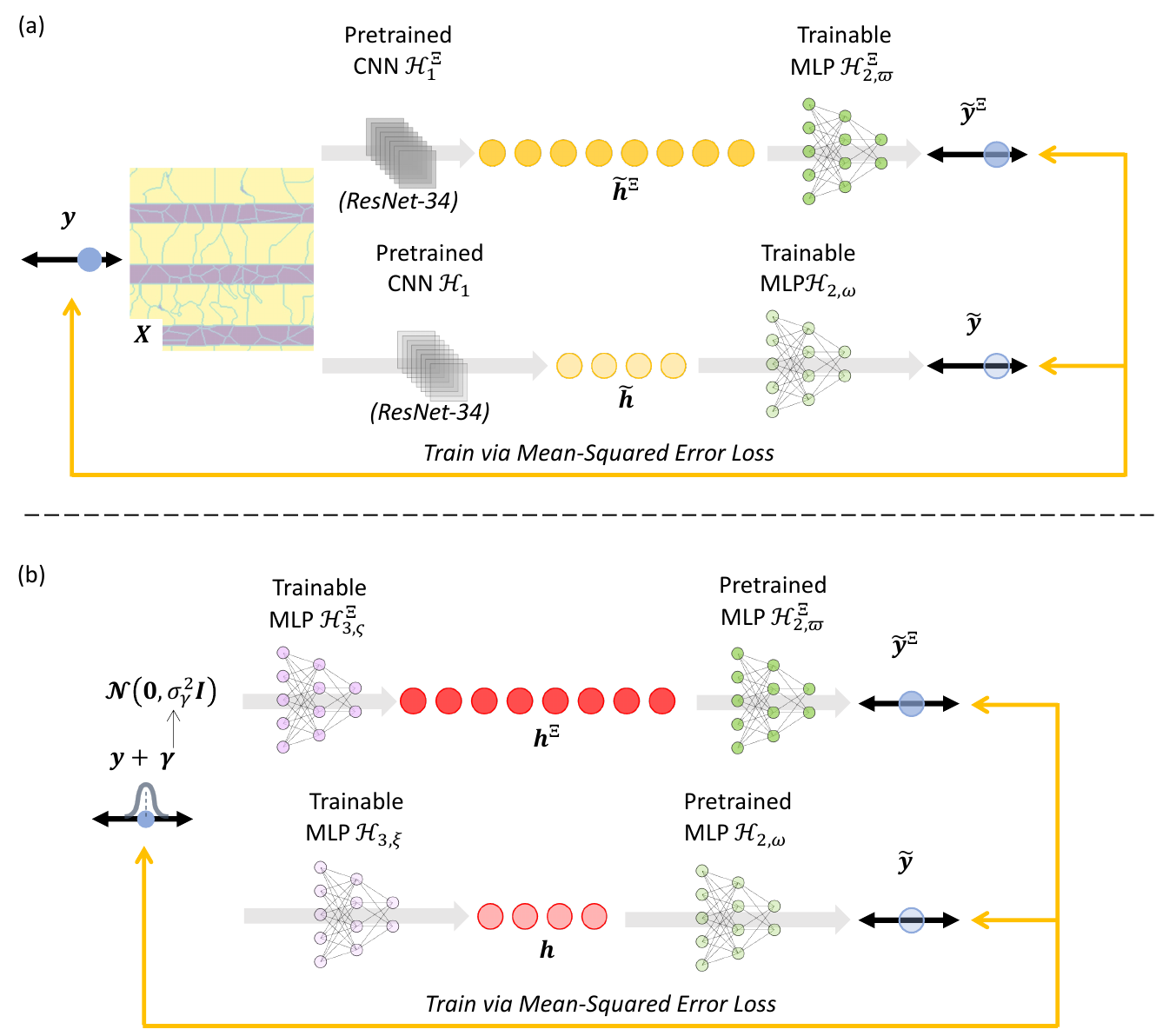}
\caption{Schematic of training scheme for high-dimensional latent embeddings $\bfh^{\bm{\Xi}}$ and $\bfh$ where we train (a) embedding-to-condition maps $\calH^{\bm{\Xi}}_{2,\varpi}$ and $\calH_{2,\omega}$ (modeled as MLPs) using pre-trained convolution NNs (ResNet-34) $\calH^{\bm{\Xi}}_1$ and $\calH_1$ and then train (b) condition-to-embedding maps $\calH^{\bm{\Xi}}_{3,\varsigma}$ and $\calH_{3,\xi}$ (modeled as MLPs) using pre-trained embedding-condition maps $\calH^{\bm{\Xi}}_{2,\varpi}$ and $\calH_{2,\omega}$ for covariance condition embedding and continuous condition embedding, respectively. In (b), condition $\bfy$ is perturbed with Gaussian noise $\bfgamma\sim\calN(\bm{0},\sigma_\gamma^2\bfI)$.}\label{fig:condition_embedding}
\end{figure}

We follow the improved label input strategy to train for condition-to-embedding maps for both covariance and continuous condition embeddings using multi-layer perceptrons (MLPs) \citep{Haykin1994}. To this end, we use a three-step process---as described by \cite{Ding2020, Ding2024}---to transform $\bfy$ into high-dimensional latent embeddings $\bfh^{\bm{\Xi}}$ and $\bfh$; see Figure \ref{fig:condition_embedding} for a visual summary of the protocol. We consider a data pair $(\bfX,\bfy)$ from the dataset $\calD$. 

\begin{enumerate}
    \item \textbf{Image-to-embedding maps}: We use pre-trained convolution NNs $\calH^{\bm{\Xi}}_1:\Rset^{C\times H\times W}\rightarrow\Rset^{d^{\bm{\Xi}}}$ and $\calH_1:\Rset^{C\times H\times W}\rightarrow\Rset^{d}$, specifically ResNet-34 \citep{He2015}, to map the image $\bfX$ to latent vectors $\tilde \bfh^{\bm{\Xi}}$ and $\tilde \bfh$ respectively, i.e.,
    \be
        \tilde\bfh^{\bm{\Xi}} = \calH^{\bm{\Xi}}_1(\bfX) \quad \text{and} \quad
        \tilde\bfh = \calH_1(\bfX).
    \ee
    
    \item \textbf{Embedding-to-condition maps}: We train two MLPs $\calH_{2,\varpi}:\Rset^{d^{\bm{\Xi}}}\rightarrow\Rset^P$ and $\calH_{2,\omega}:\Rset^{d}\rightarrow\Rset^P$ that map $\tilde \bfh^{\bm{\Xi}}$ and $\tilde \bfh$ to conditions $\tilde\bfy^{\bm{\Xi}}$ and $\tilde\bfy$ respectively, i.e.,
    \be
        \tilde\bfy^{\bm{\Xi}}=\calH^{\bm{\Xi}}_{2,\varpi}(\tilde\bfh^{\bm{\Xi}}) \quad 
        \text{and} \quad
        \tilde\bfy=\calH_{2,\omega}(\tilde\bfh),
    \ee
    where $\varpi$ and $\omega$ are the sets of trainable parameters. We train $\calH^{\bm{\Xi}}_{2,\varpi}$ and $\calH_{2,\omega}$ on the data pairs $(\bfX_i,\bfy_i)\sim\calD$ as
    \be
    \begin{aligned}
        \varpi\leftarrow & \arg\min_{\varpi}\sum_{i=1}^N \|\bfy_i - \calH^{\bm{\Xi}}_{2,\varpi}(\tilde\bfh^{\bm{\Xi}}_i) \|^2 \qquad \text{with}\quad \tilde\bfh^{\bm{\Xi}}_i =\calH^{\bm{\Xi}}_1(\bfX_i) \\
        \omega\leftarrow & \arg\min_{\omega}\sum_{i=1}^N \|\bfy_i - \calH_{2,\omega}(\tilde\bfh_i) \|^2 \qquad \text{with}\quad \tilde\bfh_i =\calH_1(\bfX_i)
    \end{aligned}
    \ee

    \item \textbf{Condition-to-embedding maps}: Lastly, we use two more MLPs $\calH^{\bm{\Xi}}_{3,\varsigma}:\Rset^{P}\rightarrow\Rset^{d^{\bm{\Xi}}}$ and $\calH_{3,\xi}:\Rset^{P}\rightarrow\Rset^{d}$ that map condition $\bfy$ to latent embeddings $\bfh^{\bm{\Xi}}$ and $\bfh$ respectively, i.e.,
    \be
    \bfh^{\bm{\Xi}} = \calH^{\bm{\Xi}}_{3,\varsigma}(\bfy) \quad 
    \text{and} \quad
    \bfh = \calH_{3,\xi}(\bfy),
    \ee
    where $\varsigma$ and $\xi$ are the sets of trainable parameters. We train $\calH^{\bm{\Xi}}_{3,\varsigma}$ and $\calH_{3,\xi}$ against their corresponding pre-trained MLPs $\calH^{\bm{\Xi}}_{2,\varpi}$ and $\calH_{2,\omega}$ as follows
    \be
    \begin{aligned}
        \varsigma\leftarrow & \arg\min_\varsigma \sum_{j=1}^M \mathbb{E}_{\bfgamma \sim \calN(\bm{0}, \sigma^2_{\bfgamma}\bfI)}  \left\|
        \calH^{\bm{\Xi}}_{2,\varpi}\left(\calH^{\bm{\Xi}}_{3,\varsigma}(\bfy_j + \bfgamma)\right)- \left(\bfy_j + \bfgamma\right)
        \right\|^2 \quad \text{and} \\
        \xi\leftarrow & \arg\min_\xi \sum_{j=1}^M \mathbb{E}_{\bfgamma \sim \calN(\bm{0}, \sigma^2_{\bfgamma}\bfI)}  \left\|
        \calH_{2,\omega}\left(\calH_{3,\xi}(\bfy_j + \bfgamma)\right)- \left(\bfy_j + \bfgamma\right)
        \right\|^2, 
    \end{aligned}
    \ee
    Here, the training is performed only over the unique conditions in the dataset (for computational efficiency) and $M$ ($\leq N$) denotes the number of unique conditions. We use perturbed conditions for training -- via added Gaussian noise $\bfgamma\sim\calN(\bm{0},\sigma_\gamma^2\bfI)$. Since the unique conditions in the dataset can be sparse, especially within an infinite space, this training strategy based on perturbed conditions encourages smoothness and vicinal continuity in the latent embeddings of these conditions.
\end{enumerate}

Another input to the model $\calF_\theta(\bfX^t,t,\bfy)$ is the step $t$ of the iterative denoising process.  We use a sinusoidal positional embedding \citep{Vaswani2017} to encode $t$ as $\bfe\in \Rset^{d^t}$, where  $d^t$ is the number of dimensions of the step embedding. The embedding is given as
\be
\bfe_i = \begin{cases}
    \sin\left( t/10000^{2i/d^t} \right) & \text{for even } i \\
    \cos\left( t/10000^{2i/d^t} \right) & \text{for odd } i
\end{cases}, \qquad \text{with} \quad i\in\{1,\dots,d^t\}.
\ee

\subsection{Data-efficient model training with vicinal loss}
\label{subsec:vicinal-loss}
For an accurate denoising process, the probability $q(\bfX^{t-1}|\bfX^t,\bfX^0,\bfy)$ in \eqref{eq:q-reverse} should be close to $p_\theta(\bfX^{t-1} \mid \bfX^t, \bfy)$ in \eqref{eq:ansatz}. Therefore, we train $\calF_\theta$ by minimizing the Kullback-Leibler (KL) divergence between the two probability distributions
\be
\begin{aligned}
    \arg\min_\theta &\  D_\text{KL}\left(q(\bfX^{t-1}|\bfX^t,\bfX^0,\bfy)\ \  \| \ \ p_\theta(\bfX^{t-1} | \bfX^t, \bfy)\right)\\
    &= \arg\min_\theta  D_\text{KL} \left(\calN(\bfX^{t-1};\bfmu_q(\bfX^t,\bfX^0,\bfy),\bfXi_q(t)) \ \  \|\ \ \calN(\bfX^{t-1};\bfmu_\theta(\bfX^t, t, \bfy), \bfXi_q(t))\right)\qquad\text{(substituting \eqref{eq:q-reverse} and \eqref{eq:ansatz})}\\
    &= \arg\min_\theta c \left(\calF_\theta(\bfX^t,t,\bfy) - \bfX^0 \right)^{T}\cdot\bfH^{-1}_{y}\cdot\left(\calF_\theta(\bfX^t,t,\bfy) - \bfX^0 \right) \qquad \text{with} \qquad c=\frac{1}{2} \frac{\bar\alpha_{t-1} (1-\alpha_t)}{(1-\bar\alpha_{t-1})(1-\bar\alpha_t)},\\
    &=  \arg\min_\theta c \Bigg[\left(\calF_\theta\left(\sqrt{\bar{\alpha}_t} \bfX^0 + \sqrt{1 - \bar{\alpha}_t} \bfepsvar,t,\bfy\right) - \bfX^0 \right)^{T} {\cdot \bfH^{-1}_{y} \cdot \left(\calF_\theta\left(\sqrt{\bar{\alpha}_t} \bfX^0 + \sqrt{1 - \bar{\alpha}_t} \bfepsvar,t,\bfy\right) - \bfX^0 \right)} \Bigg]\\
    &\qquad \qquad\text{with} \qquad \bfepsvar \sim \calN(\bm{0}, \bfH_y) \qquad \text{(substituting \eqref{eq:forward-reparam}).}
\end{aligned}
\ee
 where the KL divergence between two Gaussian distributions has been simplified analytically (see \cite{Luo2022} for detailed algebraic derivation). \cite{Ho2020} demonstrated that setting $c=1$ instead of its true value provides higher sampling quality. Therefore, we set $c=1$ in the subsequent formulation. The minimization problem can be interpreted as adding noise to the ground truth $\bfX^0$ to obtain $\bfX^t$, denoising it using $\calF_\theta$, and minimizing the discrepancy between the denoised and original versions.

Minimizing the KL divergence across all denoising steps $t\in\{1,\dots,T\}$, dataset microstructure-condition pairs $(\bfX_i,\bfy_i)\in \calD$, and noise $\bfepsvar\sim\calN(\bm{0},\bfH_y)$ can be approximated by minimizing the expectation
\be\label{eq:classical-loss-v1}
\begin{aligned}
\arg\min_\theta \frac{1}{N}\sum_{i=1}^N
\mathbb{E}_{t\sim\calU\{1,T\}}  
\mathbb{E}_{\bfepsvar\sim\calN(\bm{0},\bfH_y)} 
\Bigg[ & \left(\calF_\theta\left(\sqrt{\bar{\alpha}_t} \bfX_i + \sqrt{1 - \bar{\alpha}_t} \bfepsvar,t,\bfy_i\right) - \bfX_i \right)^{T} \\
& \cdot \bfH^{-1}_{y} \cdot \left(\calF_\theta\left(\sqrt{\bar{\alpha}_t} \bfX_i + \sqrt{1 - \bar{\alpha}_t} \bfepsvar,t,\bfy_i\right) - \bfX_i \right) \Bigg],
\end{aligned}
\ee
Here, we have dropped the superscript $(\cdot)^0$ for the sake of brevity as we only refer to ground truth data (and not intermediate noisy images). $\calU\{1,T\}$ denotes uniform sampling of natural numbers between and including $1$ and $T$.

The minimization in \eqref{eq:classical-loss-v1} can be equivalently reformulated as
\be\label{eq:classical-loss-v2}
\begin{aligned}
\arg\min_\theta \frac{1}{N}\sum_{i=1}^N
\mathbb{E}_{\bfy\sim p(\bfy)}  
\mathbb{E}_{t\sim\calU\{1,T\}}  
\mathbb{E}_{\bfepsvar\sim\calN(\bm{0},\bfH_y)} 
\Bigg[ & \mathds{1}_{\{\bfy_i=\bfy\}} 
\left(\calF_\theta\left(\sqrt{\bar{\alpha}_t} \bfX_i + \sqrt{1 - \bar{\alpha}_t} \bfepsvar,t,\bfy\right) - \bfX_i \right)^{T} \\
& \cdot \bfH^{-1}_{y} \cdot \left(\calF_\theta\left(\sqrt{\bar{\alpha}_t} \bfX_i + \sqrt{1 - \bar{\alpha}_t} \bfepsvar,t,\bfy\right) - \bfX_i \right) \Bigg],
\end{aligned}
\ee
where $\mathds{1}$ is an indicator function and $p(\bfy)$ represents the distribution of the \textit{unique} conditions in the dataset $\calD$. The optimization is carried out as an expectation over all unique conditions $\bfy$, with the indicator function ensuring that the loss for a microstructure $\bfX_i$ is activated only when its corresponding condition $\bfy_i = \bfy$ is present in the dataset.

This training approach is effective when the dataset is large enough and the continuous space of conditions $\bfy$ is sampled with adequate density. However, in low-data regimes, sparse or non-uniform sampling of the condition space fails to provide sufficient support for model generalization to unseen conditions. 

\begin{figure}
\centering
\includegraphics[width=\textwidth, trim=0cm 0cm 0cm 0cm, clip]{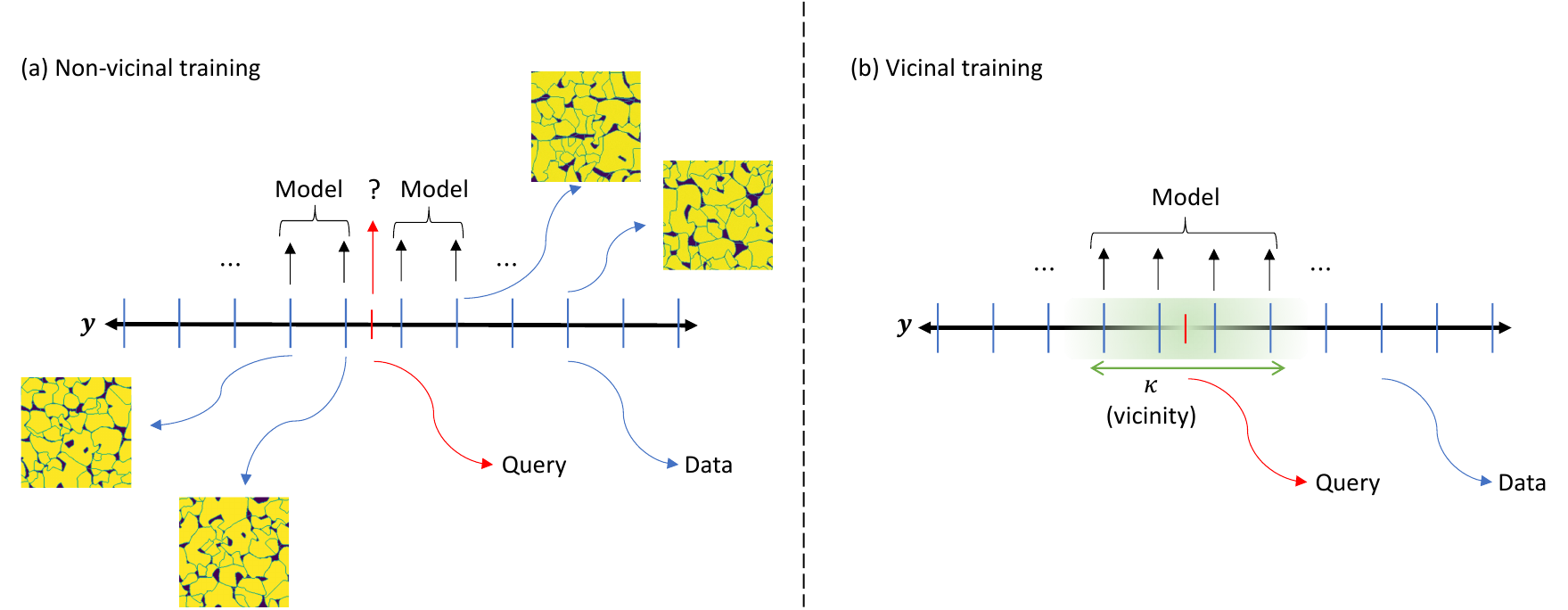}
\caption{Schematic of vicinal loss-based training. (a) In non-vicinal training, a queried process condition $\bfy$ that does not exist in the dataset is not considered during training. (b) In vicinal-loss-based training, a neighborhood determined by $\kappa$ around the query $\bfy$ is considered, even when $\bfy$  is not present in the dataset. The model learns from samples $\bfX_i$ associated with nearby conditions $\bfy_i$ within this vicinity.}\label{fig:vicinal_training}
\end{figure}

To address this issue for low-data regimes, we adapt the vicinal loss formulation of \cite{Ding2024,Ding2020}; see Figure \ref{fig:vicinal_training} for schematic. We relax the indicator function as follows
\be\label{eq:classical-loss-v3}
\begin{aligned}
\arg\min_\theta 
\frac{1}{N}\sum_{i=1}^N
\mathbb{E}_{\bfy\sim p(\bfy)}  
\mathbb{E}_{t\sim\calU\{1,T\}}  
\mathbb{E}_{\bfepsvar\sim\calN(\bm{0},\bfH_y)} 
\Bigg[ & \underbrace{\mathds{1}_{\{|\bfy-\bfy_i|\leq \kappa\}}}_{\text{\footnotesize relaxed} \atop \text{\footnotesize indicator}}\left(\calF_\theta\left(\sqrt{\bar{\alpha}_t} \bfX_i + \sqrt{1 - \bar{\alpha}_t} \bfepsvar,t,\bfy\right) - \bfX_i \right)^{T}\\
& \cdot \bfH^{-1}_{y} \cdot \left(\calF_\theta\left(\sqrt{\bar{\alpha}_t} \bfX_i + \sqrt{1 - \bar{\alpha}_t} \bfepsvar,t,\bfy\right) - \bfX_i \right)\Bigg],
\end{aligned}
\ee
such that conditions $\bfy$ that are not present in the dataset -- and thus did not participate in training previously -- are now associated with the microstructure $\bfX_i$ of a vicinal  condition $\bfy_i$ existing in the dataset $\calD$, provided they lie within a (hyperparameter) distance $\kappa>0$ of it. The unseen conditions $\bfy$ are sampled by placing Gaussian distributions centered around the existing conditions $\bfy_j$, i.e., a Gaussian kernel density estimate of $p(\bfy)$ as
\be\label{eq:p(y)}
p(\bfy) \propto \sum_{j=1}^N \exp\left(\frac{-\|\bfy-\bfy_j\|^2}{2\sigma_\delta^2}\right)
\ee
where $\sigma^2_\delta>0$ is a hyperparameter that controls the kernel variance. Expanding the expectation over $p(\bfy)$ in \eqref{eq:classical-loss-v3} and substituting  \eqref{eq:p(y)} yields the \textbf{vicinal loss training} described as 
\be
\begin{aligned}
&\arg\min_\theta  
\frac{1}{N}\sum_{i=1}^N
\int
\mathbb{E}_{t\sim\calU\{1,T\}}  
\mathbb{E}_{\bfepsvar\sim\calN(\bm{0},\bfH_y)} 
\Bigg[
\mathds{1}_{\{|\bfy-\bfy_i|\leq \kappa\}}
\left(\calF_\theta\left(\sqrt{\bar{\alpha}_t} \bfX_i + \sqrt{1 - \bar{\alpha}_t} \bfepsvar,t,\bfy\right) - \bfX_i \right)^{T} \\
& \qquad \qquad \qquad \qquad \qquad \qquad \qquad \qquad \cdot \bfH^{-1}_{y} \cdot \left(\calF_\theta\left(\sqrt{\bar{\alpha}_t} \bfX_i + \sqrt{1 - \bar{\alpha}_t} \bfepsvar,t,\bfy\right) - \bfX_i \right)
\Bigg]
 p(\bfy)\dd \bfy 
\\
 &=\arg\min_\theta 
\frac{1}{N}\sum_{i=1}^N
\sum_{j=1}^N
\int
\mathbb{E}_{t\sim\calU\{1,T\}}  
\mathbb{E}_{\bfepsvar\sim\calN(\bm{0},\bfH_y)} 
\Bigg[
\mathds{1}_{\{|\bfy-\bfy_i|\leq \kappa\}}
\left(\calF_\theta\left(\sqrt{\bar{\alpha}_t} \bfX_i + \sqrt{1 - \bar{\alpha}_t} \bfepsvar,t,\bfy\right) - \bfX_i \right)^{T}\\
& \qquad \qquad \qquad \qquad \qquad \qquad \qquad \qquad \qquad \cdot \bfH^{-1}_{y} \cdot \left(\calF_\theta\left(\sqrt{\bar{\alpha}_t} \bfX_i + \sqrt{1 - \bar{\alpha}_t} \bfepsvar,t,\bfy\right) - \bfX_i \right)
\Bigg]
 \exp\left(\frac{-\|\bfy-\bfy_j\|^2}{2\sigma_\delta^2}\right) \dd \bfy.
\end{aligned}
\ee

Setting $\bfdelta = \bfy-\bfy_j$ simplifies the optimization into
\be
\begin{aligned}
&
\arg\min_\theta 
\frac{1}{N}\sum_{i=1}^N
\sum_{j=1}^N
\int
\mathbb{E}_{t\sim\calU\{1,T\}}  
\mathbb{E}_{\bfepsvar\sim\calN(\bm{0},\bfH_{\bfy_j+\bfdelta})} 
\Bigg[
\mathds{1}_{\{|\bfy_j+\bfdelta-\bfy_i|\leq \kappa\}}
\left(\calF_\theta\left(\sqrt{\bar{\alpha}_t} \bfX_i + \sqrt{1 - \bar{\alpha}_t} \bfepsvar,t,\bfy_j+\bfdelta\right) - \bfX_i \right)^{T} \\
& \qquad \qquad \qquad \qquad \qquad \qquad \qquad \qquad \qquad \cdot \bfH^{-1}_{\bfy_j+\bfdelta} \cdot \left(\calF_\theta\left(\sqrt{\bar{\alpha}_t} \bfX_i + \sqrt{1 - \bar{\alpha}_t} \bfepsvar,t,\bfy_j+\bfdelta\right) - \bfX_i \right)
\Bigg]
 \exp\left(\frac{-\|\bfdelta\|^2}{2\sigma_\delta^2}\right) \dd \bfdelta
\\
&=
\arg\min_\theta 
\frac{1}{N}\sum_{i=1}^N
\sum_{j=1}^N
\mathbb{E}_{\bfdelta\sim\calN(\boldsymbol{0},\sigma_\delta^2\bfI)}
\mathbb{E}_{t\sim\calU\{1,T\}}  
\mathbb{E}_{\bfepsvar\sim\calN(\bm{0},\bfH_{\bfy_j+\bfdelta})} 
\Bigg[
\mathds{1}_{\{|\bfy_j+\bfdelta-\bfy_i|\leq \kappa\}}
\left(\calF_\theta\left(\sqrt{\bar{\alpha}_t} \bfX_i + \sqrt{1 - \bar{\alpha}_t} \bfepsvar,t,\bfy_j+\bfdelta\right) - \bfX_i \right)^{T}\\
& \qquad \qquad \qquad \qquad \qquad \qquad \qquad \qquad \qquad \qquad \qquad \cdot \bfH^{-1}_{\bfy_j+\bfdelta} \cdot \left(\calF_\theta\left(\sqrt{\bar{\alpha}_t} \bfX_i + \sqrt{1 - \bar{\alpha}_t} \bfepsvar,t,\bfy_j+\bfdelta\right) - \bfX_i \right)
\Bigg].
\end{aligned}
\label{eq:vicinal_loss}
\ee
In summary, the vicinal loss makes the strict condition–microstructure associations in the dataset fuzzy during training, thereby promoting a smoother mapping from conditions to microstructures, especially in data-sparse regions of the continuous condition space.

\subsection{Model architecture}
\label{subsec:model_arch}

\begin{figure}[H]
\centering
\includegraphics[width=0.8\textwidth]{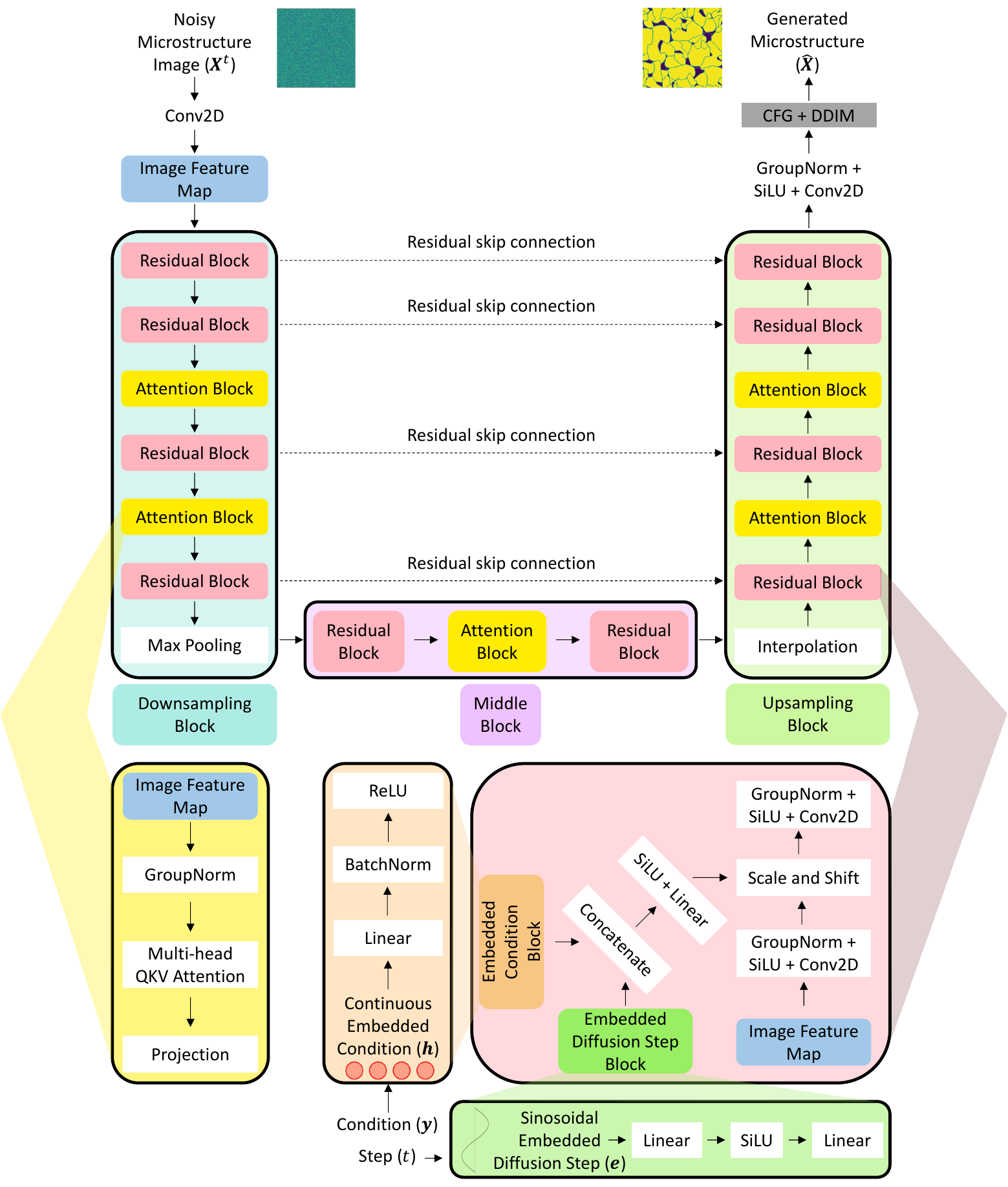}
\caption{U-net network architecture of the denoising diffusion model, including  CFG and DDIM, for generation with continuous conditioning.}
\label{fig:unet}
\end{figure}

We use a U-net architecture proposed by \cite{Ronneberger2015} to model $\calF_\theta(\bfX^t,t,\bfy)$; see Figure \ref{fig:unet} for schematic. The U-net consists of two \textbf{embedding blocks} and multiple layers of \textbf{residual blocks} and \textbf{self-attention blocks}, which sequentially transform the input image $\bfX^t$ into lower-dimensional representations (\textbf{downsampling block}), then having a set of residual and attention blocks preserving the image dimensions (\textbf{middle block}), and then reconstruct it back (\textbf{upsampling block}) to an output with the same dimensions as $\bfX^t$.

\textbf{Embedding blocks:} The continuous condition and denoising step embeddings ($\bfh$ and $\bfe$, respectively) are first processed by {embedding blocks} composed of MLPs. For the embedded diffusion step block, we have a linear layer \citep{Moczulski2015}, followed by a SiLU activation layer \citep{Elfwing2018} and a second linear layer. For the embedded condition block, we have a linear layer, followed by batch normalization \citep{Ioffe2015}, and then finally a ReLU activation layer \citep{Agarap2018}. The resulting outputs are then fed into each residual block throughout the network.

\textbf{Residual blocks:} Inside each residual block, the input image is processed by a \textbf{Conv2D block}  that contains sequentially group normalization \citep{Wu2018}, SiLU activation and a convolution layer \citep{Oshea2015}. In parallel, the condition and step embedding (outputs of the embedding blocks) are concatenated and processed by a SiLU activation and a linear layer. The processed embedding are split into two chunks, each of which is used to scale and shift the outputs of the convolution layers and modulate the normalized feature map. Finally, another Conv2D block is applied and the output is passed to the next block.

During the upsampling stage of the U-net, skip connections \citep{Wu2020} channel-wise concatenate the intermediate outputs from the downsampling residual blocks with the inputs to the corresponding upsampling residual blocks.

\textbf{Attention blocks:} Inside each attention block, we use group normalization followed by multi-head self-attention with query-key-value structure \citep{Vaswani2017} to extract long-range structure and global context in the image, and acquire and add the projection with the input feature map, before being passed to the next block. Note that we do not use skip connections with attention blocks.

The final operation in the downsampling block involves the process of applying a convolution layer or max pooling \citep{Matoba2022} before moving onto the middle block. Similarly, the initial operation after receiving the feature map at the end of the middle block involves interpolation using the nearest-neighbor mode \citep{Rukundo2019} which increases the spatial resolution of the feature map before proceeding to the residual and attention blocks.

 After the final residual block operation in the upsampling block, the image goes through the final Conv2D block, which includes a group normalization, SiLU activation and convolution layer, to restore the dimensions of the {U-net output $\calF_\theta(\bfX^t,t,\bfy)$} back to the dimensions of the original U-net input $\bfX^t$. During training, we wrap U-net model using the \texttt{accelerate} python package \citep{accelerate} for multi-GPU training and exponential moving average \citep{Morales-Brotons2024} to stabilize training and convergence.

{In the following subsection, we describe the sampling strategy employed to transform $\calF_\theta(\bfX^t,t,\bfy)$ into the generated microstructure image $\hat{\bfX}$.} 

\subsection{High-quality and efficient sampling with CFG and DDIM}
\label{subsec:cfg_ddim}

To ensure high-quality microstructure generation, we use CFG \citep{Ho2022} to promote conditional generation.

We train the U-net simultaneously as both conditional and unconditional models using a hyperparameter called dropout probability $p_{\text{drop}} \in [0,1]$. At each training iteration, we sample a random value $u$ from a uniform distribution $\calU(0, 1)$. If $u \geq p_{\text{drop}}$, we provide the latent condition embedding $\bfh$ to the model and conditionally train it; {otherwise, we replace the condition embedding with a null embedding $\emptyset$. We initialize $\emptyset$ as a trainable parameter sampled from a negatively valued random vector. The U-net, therefore, will distinguish unconditional embedding from the conditional counterpart based on the sign.}

After performing this joint training, we replace the  U-net output $\calF_\theta(\bfX^t,t,\bfy)$ with $\calF_\theta^\text{CFG}(\bfX^t,t,\bfy)$ given by:
\be\label{eq:CFG}
\calF_\theta^\text{CFG}(\bfX^t,t,\bfy) = \underbrace{\calF_\theta(\bfX^t,t,\emptyset)}_{\text{\footnotesize unconditional prediction}} + \quad 
\Gamma \underbrace{  (\calF_\theta(\bfX^t,t,\bfy) - \calF_\theta(\bfX^t,t,\emptyset))}_{\text{\footnotesize conditional guidance}}.
\ee 
Here, the \textit{unconditional prediction}, defined as the U-net output with a null embedding as input, represents the general distribution of all microstructures. In contrast, the \textit{conditional guidance} term provides an implicit bias toward the distribution of microstructures corresponding to a specific condition $\bfy$ and away from the unconditional distribution. The parameter $\Gamma > 0$ is a scaling hyperparameter that controls the relative strength of conditional guidance over the unconditional prediction. When $\Gamma = 0$, the model output reduces to unconditional generation, whereas $\Gamma = 1$ corresponds to conditional generation with no CFG. In practice, $\Gamma > 1$ is often used to encourage stronger alignment with the conditional distribution, although this typically comes at the cost of reduced diversity in the generated microstructures. We also apply more advanced guidance refinements like orthogonal decomposition and output rescaling, however for clarity, we present only the core formulation in \eqref{eq:CFG} (see \cite{Ding2024} and \citet{Wang2024DDPM} for detailed implementation).

The CFG output $\calF_\theta^\text{CFG}(\bfX^t,t,\bfy)$ (shortened to $\calF_\theta^\text{CFG}$ for brevity) simply replaces $\calF_\theta(\bfX^t,t,\bfy)$ in the mean $\bfmu_\theta(\bfX^t,t,\bfy)$ \eqref{eq:ansatz_mean} of the posterior distribution $p_\theta(\bfX^{t-1} \mid \bfX^t, \bfy)$ \eqref{eq:ansatz}. Therefore, we obtain the stochastic denoising process proposed by \cite{Ho2020} as 
\be\label{eq:DDPM_sampling}
\bfX^{t-1} = \sqrt{\bar{\alpha}_{t-1}} \calF_\theta^\text{CFG} + \sqrt{1 - \bar{\alpha}_{t-1} - \sigma_t^2} \frac{\bfX^{t} - \sqrt{\bar{\alpha}_t} \calF_\theta^\text{CFG}}{\sqrt{1 - \bar{\alpha}_t}} + \sigma_t \bfepsvar \quad \text{with} \quad \bfepsvar \sim \calN(0, \bfI), \quad \bar{\alpha}_t = \prod_{i=1}^t \alpha_i,\quad \text{and} \quad \alpha_t = 1 - \beta_t
\ee
where $\bfX^t$ is the noisy microstructure image at step $t\in\{1,\dots,T\}$, $\bfX^T$ is $\bfy$-dependent Gaussian noise $\bfX^T\sim\calN(\bm{0},\bfH_y)$, and $\sigma_t$ controls the stochastic variance to the amount of noise added during sampling, defined as
\be
\sigma_t = \nu \cdot \sqrt{\frac{(1 - \frac{\alpha_t}{\alpha_{t-1}}) \cdot (1 - \alpha_{t-1})}{(1 - \alpha_t)}}
\ee
where $\nu = 1$ is a constant for a probabilistic denoising process.

However, \cite{Song2020} propose DDIM to expedite sampling by setting the value of $\nu = 0$ and making the denoising process deterministic. By cutting the denoising steps from $T$ to $T' \ll T$, DDIM preserves sample consistency, i.e., images derived from similar latent variables share comparable high-level features. DDIM modifies  \eqref{eq:DDPM_sampling} as
\be\label{eq:DDIM_sampling}
\bfX^{t-1} = \sqrt{\bar{\alpha}_{t-1}} \calF_\theta^\text{CFG} + \sqrt{1 - \bar{\alpha}_{t-1}} \frac{\bfX^{t} - \sqrt{\bar{\alpha}_t} \calF_\theta^\text{CFG}}{\sqrt{1 - \bar{\alpha}_t}} \quad \text{with}\quad \bar{\alpha}_t = \prod_{i=1}^t \alpha_i,\quad \text{and}\quad \alpha_t = 1 - \beta_t
\ee

where $\bfX^t$ is the noisy microstructure image at step $t\in\{1,\dots,T'\}$ provided as input to the U-net. {We obtain the generated microstructure image $\hat{\bfX}$ after iterating till the last denoising step.}

An overview of the model architecture including the sampling strategy block is provided in Figure \ref{fig:unet}.
All hyperparameter values used for training and sampling are summarized in \ref{sec:parameters}.

\section{Results}
\label{sec:results}

Here, we evaluate the performance of the proposed generative ML framework for conditional microstructure generation. We select 10 equally spaced manganese concentration values, $\chi^\text{avg}_\text{Mn}$, within the range of 0.0021 to 0.0220 atomic fraction, ensuring that these values are not included in the training dataset. For each unseen condition, we generate 100 microstructure images using both the diffusion model and ground-truth simulations and then perform qualitative  (Section \ref{subsubsec:qual_results_best}) and quantitative comparisons (Section \ref{subsec:quant_results_best}). Note that, in the context of generative modeling, the generated images are not expected to match the ground truth exactly, but rather to reproduce the underlying statistical and morphological characteristics.

Additionally, we perform \textbf{ablation studies} by selectively disabling three key training strategies---introduced in previous sections---to assess their impact on model performance. Namely,
\begin{itemize}
    \item effect of \textbf{denoising diffusion implicit method (DDIM)} for efficient microstructure image sampling (Section \ref{subsec:ddim_vs_no_ddim}),
    \item effect of \textbf{ classifier-free guidance (CFG)} for higher quality microstructure image sampling (Section \ref{subsubsec:cfg_vs_no_cfg}),
    \item effect of \textbf{vicinity-based loss} for data-efficient model training (Section \ref{subsubsec:vicinity_vs_no_vicinity}), and
    \item effect of \textbf{all of the above combined} (Section \ref{subsubsec:all_on_vs_all_off}).
\end{itemize}

For clarity, each results figure indicates at the top which of the three strategies (DDIM, CFG, and vicinity-based loss) are enabled or disabled. The ``\textbf{diffusion model}'' refers to the configuration in which all three strategies are enabled, whereas the ``\textbf{ablation model}'' refers to configurations in which one or more strategies are disabled.

\subsection{Diffusion model: qualitative microstructure analysis}
\label{subsubsec:qual_results_best}

First, we apply thresholding to the microstructure images generated by the diffusion model to ensure clear separation of pixel values as defined in the ground truth microstructure image representation (see Section \ref{subsec:problem_setting}). The thresholding assigns $-1$ to austenite grain interiors, $-0.25$ to austenite boundaries, $0.25$ to ferrite boundaries, and $1$ to ferrite grain interiors. Therefore, for a normalized pixel value $p_{\text{model}}\in[-1,1]$ in the microstructure image generated by the diffusion model, we assign the post-processed pixel value $p$ as

\be
\begin{cases}
    \text{Austenite grain interior} &\implies p = -1, \quad \text{if} \quad -1 \leq p_{\text{model}} < -0.3 \\
    \text{Austenite boundary} &\implies p = -0.25, \quad \text{if} \quad -0.3 \leq p_{\text{model}} < 0 \\
    \text{Ferrite boundary} &\implies p = 0.25, \quad \text{if} \quad 0 \leq p_{\text{model}} < 0.3 \\
    \text{Ferrite grain interior} &\implies p = 1, \quad \text{if} \quad 0.3 \leq p_{\text{model}} \leq 1 
\end{cases}
\ee

Further, the thresholded microstructure images are processed using the connected-component labeling algorithm from open-source software \texttt{Scipy} \citep{Virtanen2020} to identify contiguous grain regions and produce grain morphology maps. 
We simultaneously acquire phase maps by interpreting the signed pixel values---negative pixel values correspond to austenite phase and positive pixel values correspond to ferrite phase. We further compute two-point autocorrelation of the phase maps for qualitative comparison.

\begin{figure}[H]
\centering
\includegraphics[width=0.8\textwidth]{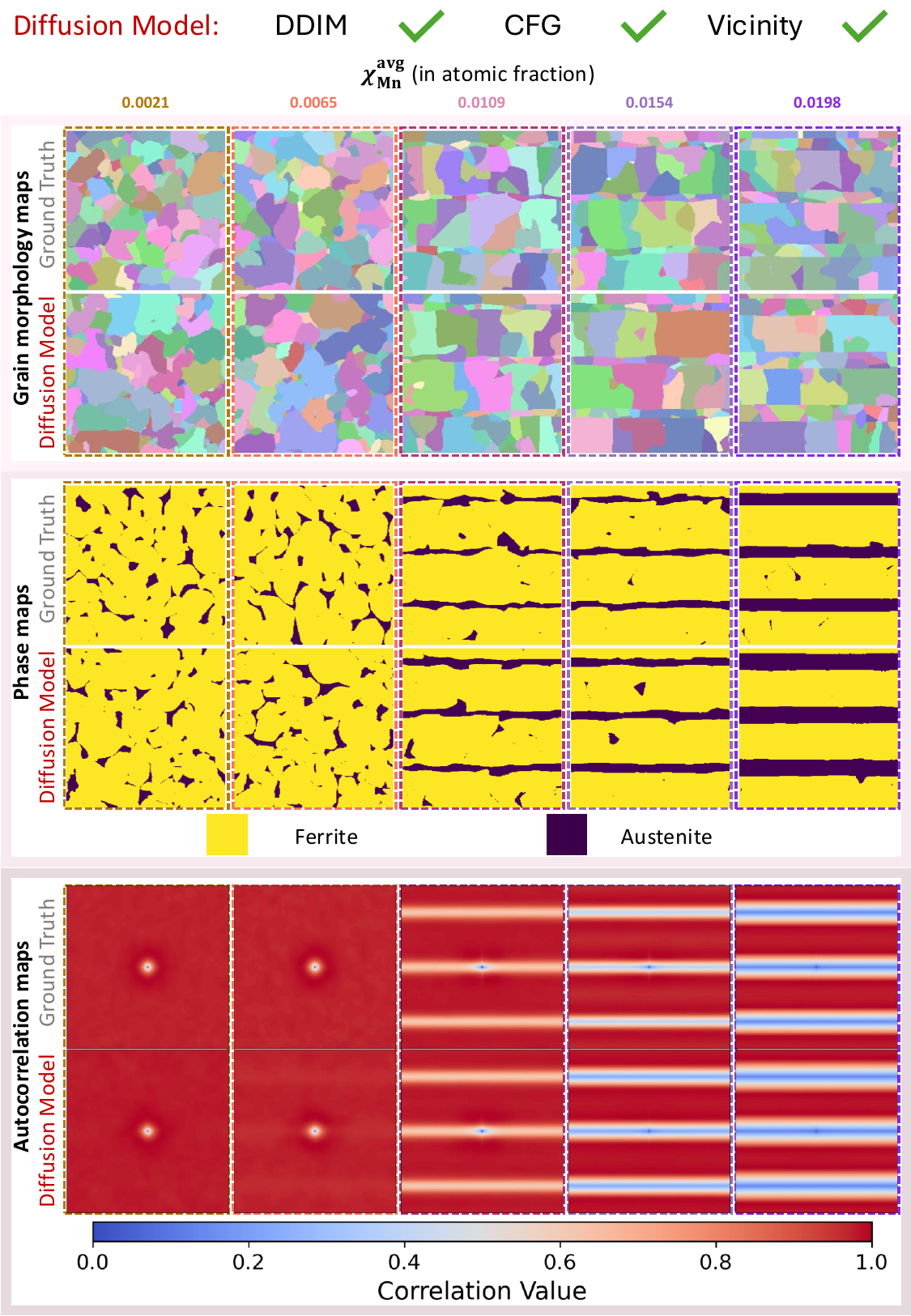}
\caption{Qualitative comparison of microstructures between ground truth and diffusion model at specific $\chi^\text{avg}_\text{Mn}$ values (indicated above the top panel) not present in the training dataset. \textit{Top panel:} grain morphology maps. \textit{Middle panel:} phase maps. \textit{Bottom panel:} autocorrelation maps. The diffusion model produces microstructures that are in qualitative agreement with the ground truth.}
\label{fig:combined_qual_best}
\end{figure}

In Figure \ref{fig:combined_qual_best}, the representative grain morphology and phase maps generated from the diffusion model show good agreement with representative ground truths. The model demonstrates the ability to interpolate meaningful and high-fidelity microstructure representations in latent space despite not being trained on the specific average manganese concentration values ($\chi^\text{avg}_\text{Mn}$) presented here. The reconstruction of austenite bands is particularly accurate at higher $\chi^\text{avg}_\text{Mn}$---consistent with the fact that manganese acts as an austenite stabilizer that suppresses the driving force for austenite-to-ferrite phase transformation \citep{Bandi2021}.

We also plot the autocorrelation maps between the generated phase maps by the diffusion model and ground truth in Figure \ref{fig:combined_qual_best} for quantitative comparison of  the spatial arrangement of the phases. The ground truth exhibits pronounced anisotropy in the correlation map at higher $\chi^\text{avg}_\text{Mn}$, which is also reproduced accurately by the diffusion model. Additionally, the similarity between the autocorrelation maps implies that the diffusion model has successfully learned not only the local morphology but also the long-range spatial patterns governed by the manganese distribution described in Section \ref{subsec:data_generation}. 

\begin{figure}[H]
\centering
\includegraphics[width=\textwidth]{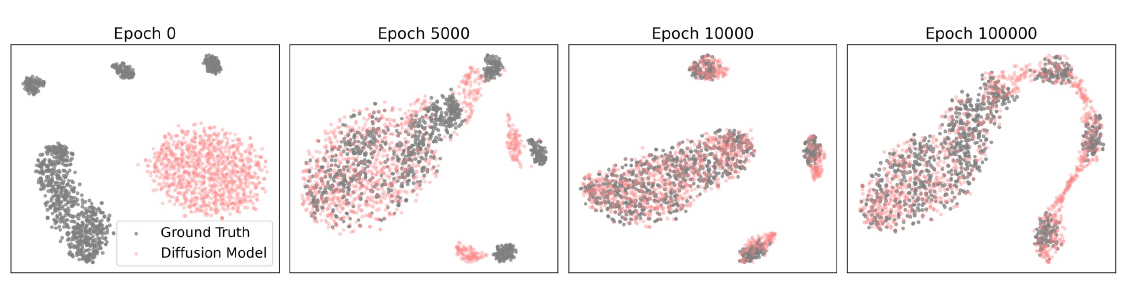}
\caption{Evolution of the t-SNE map of ground-truth and generated microstructures at various training epochs.}
\label{fig:tsne}
\end{figure}

The above observation is further supported by the t-distributed stochastic neighbor embedding (t-SNE) \citep{vandermaaten2008} of both generated and ground truth microstructures at different stages of training, as shown in Figure \ref{fig:tsne}. At the start of training, i.e.,  at the zeroth epoch, the diffusion model outputs are essentially noise, and their corresponding embeddings lie far from those of the ground truth microstructures. As training progresses from $0$ to $5{,}000$ epochs and then to $10{,}000$ epochs, the model gradually learns to capture key microstructural features. This is reflected in the migration of the generated image embeddings toward those of the ground truth. By the end of training, the embeddings overlap, indicating that the generated and ground truth distributions become closely aligned after $100{,}000$ epochs. Interestingly, the embeddings from the diffusion model interpolate between distinct clusters in the ground truth data, indicating that the model can learn meaningful representations even in regions where labeled data is absent or sparse. \ref{sec:loss_curves} further provides the training and validation loss curves of the diffusion model.

\subsection{Diffusion model: Quantitative microstructure analysis}
\label{subsec:quant_results_best}

\begin{figure}
\centering
\includegraphics[width=0.83\textwidth]{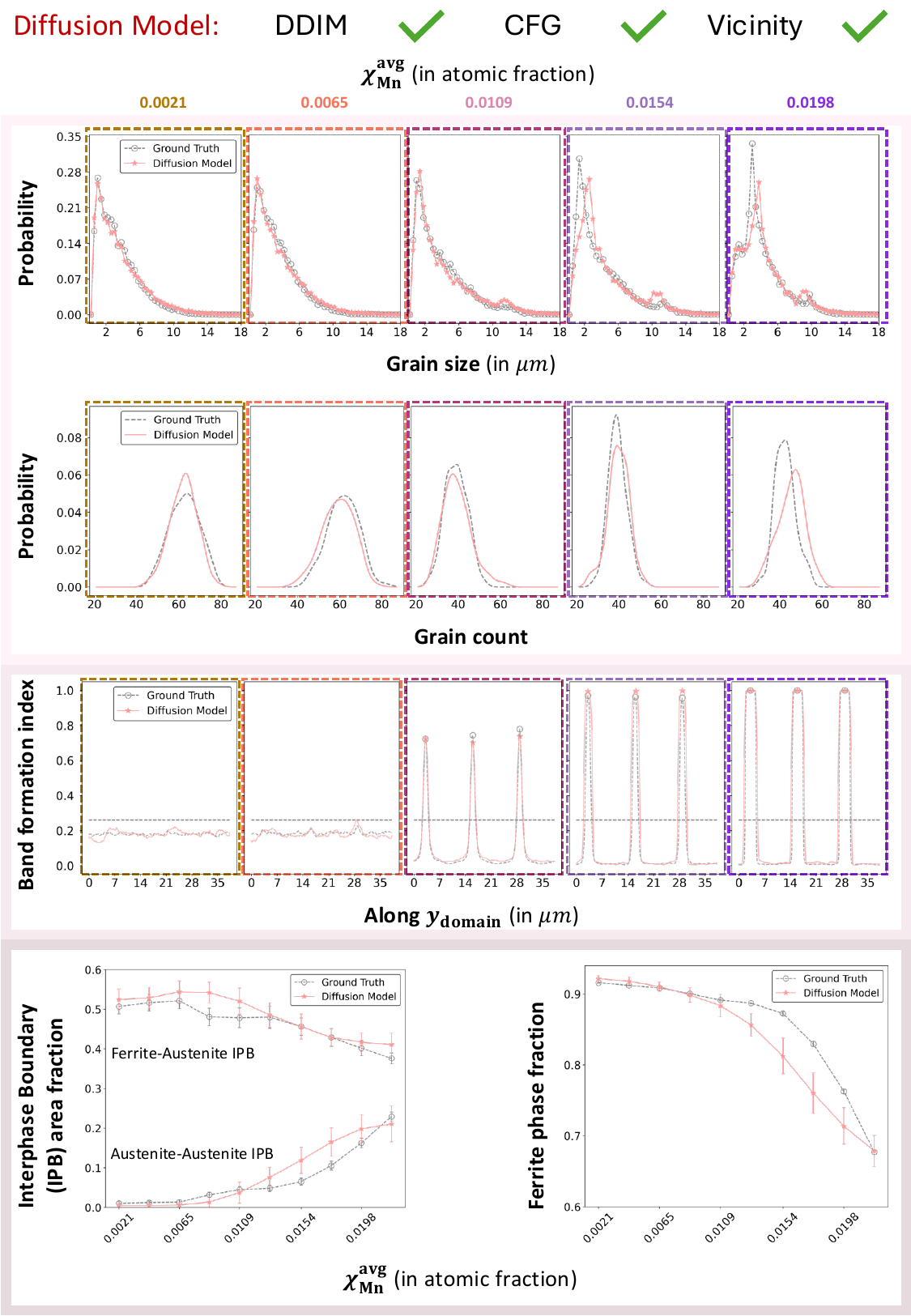}
\caption{Quantitative comparison of microstructure statistics between ground truth (gray dashed line with circles) and diffusion model (red line with filled stars) at specific $\chi^\text{avg}_\text{Mn}$ not present in the training dataset. The $\chi^\text{avg}_\text{Mn}$ values for the top and middle panels are indicated above the top panel, whereas in the bottom panel they are shown along the horizontal axes of the respective plots. \textit{Top panel:} ferrite grain size and ferrite grain count distributions.  \textit{Middle panel:}  band formation index along $y_{\text{domain}}$, i.e., the y-direction inside the microstructure domain. \textit{Bottom panel:} ferrite-austenite and austenite-austenite IPB area fractions and ferrite phase fraction vs.~$\chi^\text{avg}_\text{Mn}$. The statistics of the microstructures predicted by the diffusion model are in good agreement with those of the ground truth.}
\label{fig:combined_quant_best}
\end{figure}

To evaluate the performance of the diffusion model, we use quantitative metrics as qualitative assessments of microstructures offer limited insights, and manual inspection of large datasets is impractical. We use a range of descriptors to quantify our microstructures including the grain size distribution, grain count distribution, band formation index, IPB area fractions, and phase fraction; see Figure \ref{fig:combined_quant_best}. Quantitative statistics were averaged across 100 microstructure images per condition value chosen. While phase fractions are straightforward, the remaining descriptors and their computation are briefly discussed here.

The grain size is determined using the intercept method, following the ASTM E112 standard for two-phase microstructures \citep{ASTM_E112}. This approach involves drawing randomly oriented lines across the microstructure and counting their intersections with grain boundaries. The lengths of the line segments within individual grains and phases are measured, and the process is repeated across multiple lines and images to obtain the intercept length distribution.

Grain count distributions are obtained by fitting a kernel density estimation (KDE) to the grain counts per image. To prevent small-pixel artifacts from being counted as grains, only grains with an area-equivalent diameter exceeding 4 pixels are considered (see Section \ref{subsec:limitations}).

We use band formation index \citep{Farahani2018, Gutirrez2016} to quantify the austenite banding effect due to the spatially-varying presence of manganese. Band formation index is calculated as the fraction of austenite phase for each horizontal row of pixels in the microstructure along $y_{\text{domain}}$ ($y$-direction inside the microstructure domain). Peaks in band formation index correspond to regions with a high austenite fraction and, correspondingly, high manganese concentration.

The IPB area fractions are computed for three boundary types: ferrite-austenite, austenite-austenite, and ferrite–ferrite  boundaries. These values are obtained by quantifying the total pixel count for each boundary type in the microstructure images and expressing them as fractional proportions. Since the sum of all three IPB area fractions equals 1, ferrite–ferrite IPB area fractions are omitted from the analysis to avoid redundancy.

For the grain size distribution, grain count distribution, and band formation index, we analyze the generated microstructures for five representative and unseen cases in Figure \ref{fig:combined_quant_best}. The ferrite grain size distribution from the diffusion model closely matches that of the ground truth, with only minor deviations at higher manganese content values. This may be attributed to challenges in resolving open boundaries, which can lead to inadequate grain identification (see Section \ref{subsec:limitations}). The KDE fit to the ferrite grain count distributions also shows good agreement, except at higher manganese content values---again attributable to difficulties in open grain boundary generation. We highlight that both grain size and count are stochastic quantities whose accuracy cannot be evaluated discriminately (i.e., there is no true value). The agreement between the true and predicted distributions (via KDE fits) support the assertion that the diffusion model captures the stochastic nature inherent to our ground truth microstructure images. We also observe good agreement in the band formation index, with the diffusion model accurately capturing the onset of banding as the manganese concentration increases.

For IPB area fractions and phase fractions, we analyze 10 equally spaced, previously unseen values of $\chi^\text{avg}_\text{Mn}$ between $0.0021$ and $0.0220$ (see Figure \ref{fig:combined_quant_best}). As the manganese concentration increases, austenite becomes more stable, leading to an increase in austenite–austenite grain boundary formation and a corresponding decrease in ferrite–austenite interfacial boundaries. The diffusion model reproduces this trend in the IPB area fractions and shows good agreement with the ground truth. Similar to band formation index, limitations in resolving open boundaries in the generated microstructures during post-processing may introduce discrepancies in the IPB area fractions (see Section \ref{subsec:limitations}).

Lastly, the ferrite phase fraction shows good agreement with the ground truth at lower manganese concentrations, but deviations are observed at higher concentrations. This may be attributed to slight differences in austenite band widths at higher manganese concentrations between the diffusion model and the ground truth, which reduce the total number of ferrite regions in the former and is consistent with the larger error bars observed.

While the diffusion model shows good qualitative and quantitative agreement with the ground truth, we now conduct ablation studies to assess the roles of different training and sampling strategies in overall microstructure generation.

\subsection{Ablation study: DDIM}
\label{subsec:ddim_vs_no_ddim}

\begin{figure}[H]
\centering
\includegraphics[width=0.83\textwidth]{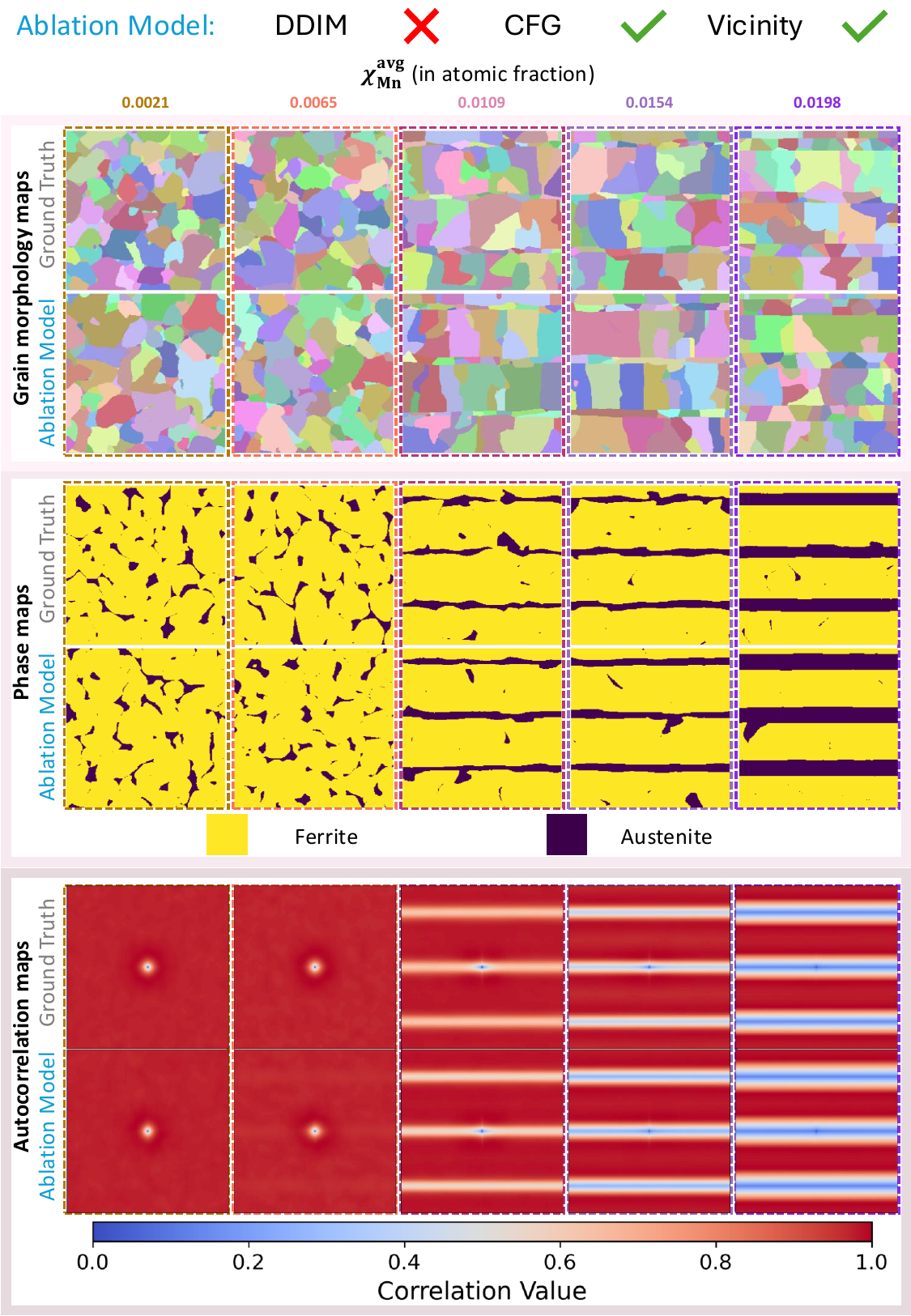}
\caption{\textbf{DDIM ablation}: Qualitative comparison of microstructures between ground truth and DDIM-ablation model at specific $\chi^\text{avg}_\text{Mn}$ values (indicated above the top panel) not present in the training dataset. \textit{Top panel:} grain morphology maps. \textit{Middle panel:} phase maps. \textit{Bottom panel:} autocorrelation maps. The DDIM-ablation model still produces microstructures that are in qualitative agreement with the ground truth.}
\label{fig:combined_qual_no_DDIM}
\end{figure}

\begin{figure}[H]
\centering
\includegraphics[width=0.83\textwidth]{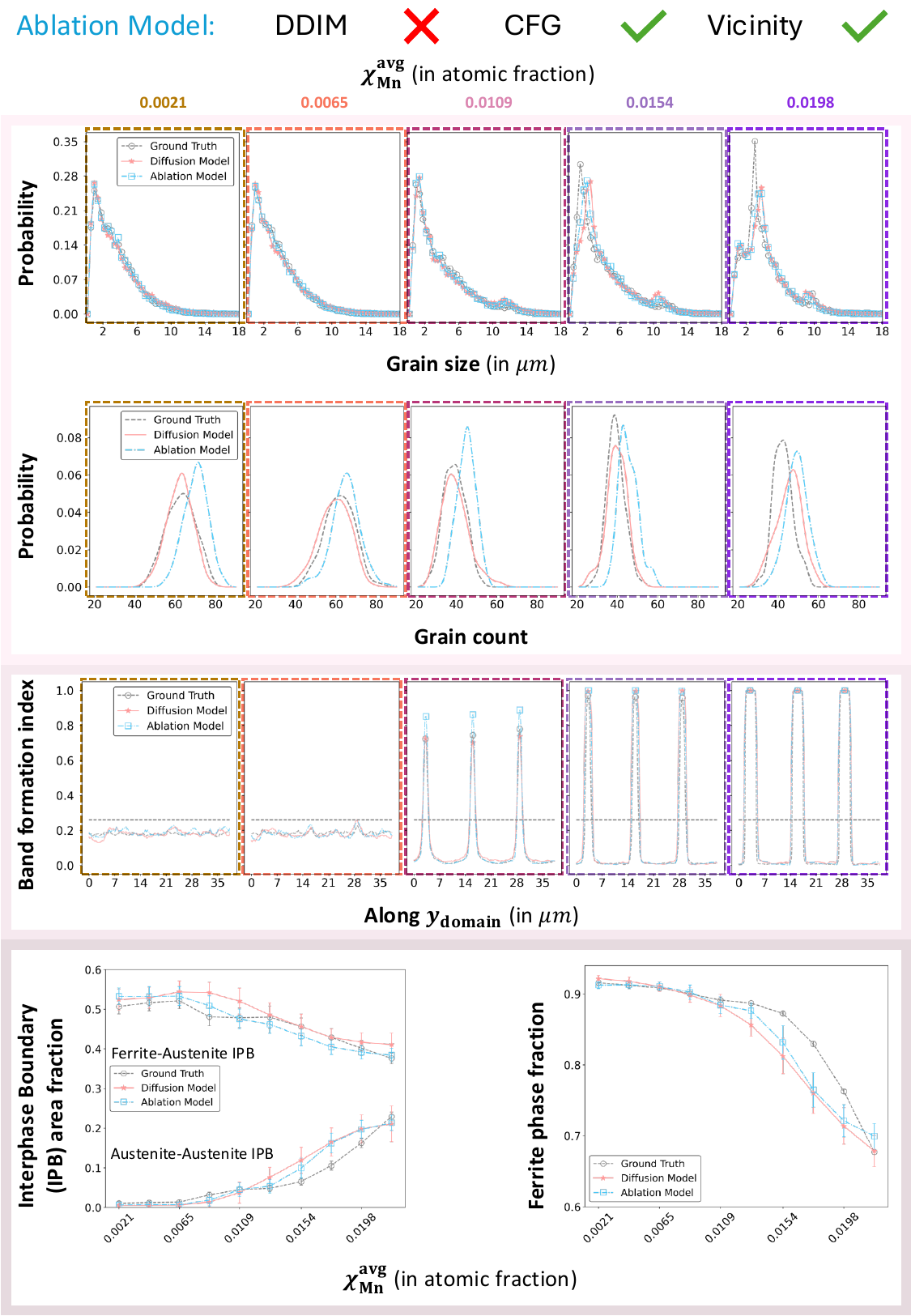}
\caption{\textbf{DDIM ablation:} Quantitative comparison of microstructure statistics between ground truth (gray dashed line with circles) and DDIM-ablation model (red line with filled stars) at specific $\chi^\text{avg}_\text{Mn}$ not present in the training dataset. The $\chi^\text{avg}_\text{Mn}$ values for the top and middle panels are indicated above the top panel, whereas in the bottom panel they are shown along the horizontal axes of the respective plots. \textit{Top panel:} ferrite grain size and ferrite grain count distributions.  \textit{Middle panel:}  band formation index along $y_{\text{domain}}$, i.e., the y-direction inside the microstructure domain. \textit{Bottom panel:} ferrite-austenite and austenite-austenite IPB area fractions and ferrite phase fraction vs.~$\chi^\text{avg}_\text{Mn}$. Key differences are observed in band formation index and ferrite grain count distribution.}
\label{fig:combined_quant_no_DDIM}
\end{figure}

We assess an \textbf{ablation model in which accelerated sampling via DDIM in \eqref{eq:DDIM_sampling} is disabled and replaced with standard sampling in \eqref{eq:DDPM_sampling}.} Figure \ref{fig:combined_qual_no_DDIM} provides a qualitative comparison of the DDIM-disabled microstructure generation with the ground truth, while Figure \ref{fig:combined_quant_no_DDIM} presents a quantitative comparison with both the ground truth and the diffusion model.

The generated grain morphology, phase, and autocorrelation maps show good agreement with the ground truth, similar to the results obtained using DDIM sampling. However, some quantitative discrepancies are observed. In particular, the ablation model overestimates the ferrite grain counts, as indicated by the rightward shift in the means of its KDEs. Additionally, for $\chi^\text{avg}_\text{Mn} = 0.0109$, the banding effect is more pronounced in the band formation index results than in either the ground truth or the diffusion model with DDIM. The remaining quantitative metrics show comparable performance between the ablation and diffusion models, with differences generally falling within the observed variance or error bars.

\begin{figure}[H]
\centering
\includegraphics[width=0.7\textwidth]{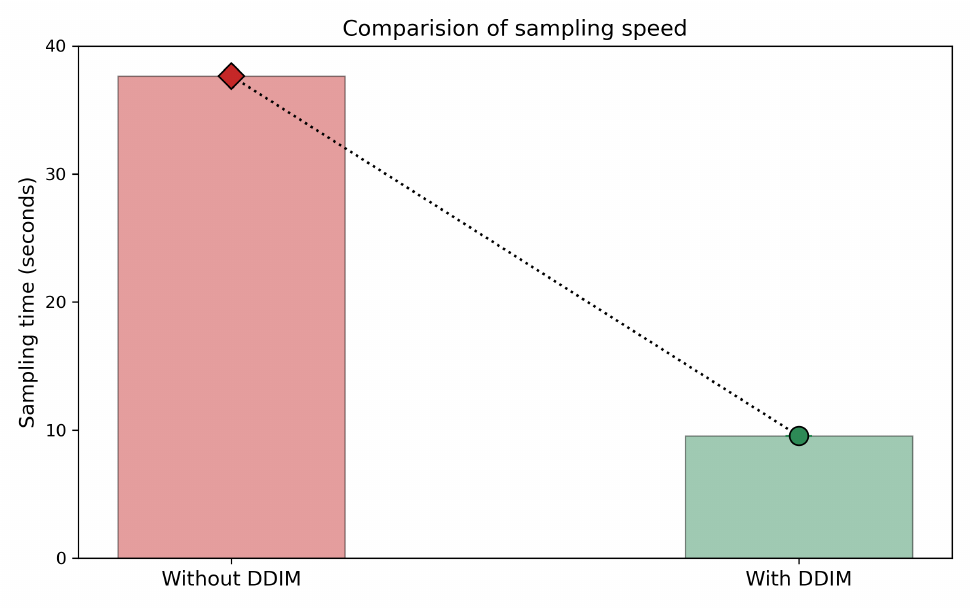}
\caption{Comparison of sampling speed (averaged across $1{,}000$ generations) between DDIM and the ablation model discussed in Section \ref{subsec:ddim_vs_no_ddim}. DDIM achieves up to a $4\times$ increase in sampling speed relative to the ablation setting where it is disabled. See \ref{sec:parameters} for hardware specifications and runtime details.}
\label{fig:sampling_strat}
\end{figure}

That said, the primary advantage of DDIM lies in its computational efficiency during sampling. Sampling speed is an important factor for practical applications of generative ML models, such as real-time process-structure-property optimization, where it offers a clear benefit over trial-and-error approaches based on physics-based models. Figure \ref{fig:sampling_strat} compares the average sampling time, showing that DDIM achieves a 4x speedup relative to the ablation model without DDIM, see \ref{sec:parameters} for hardware specifications and runtime details.

Overall, the ablation study demonstrates that DDIM provides significant acceleration in sampling without compromising qualitative microstructure fidelity, and even yields slightly improved performance in the statistical analysis of the generated microstructures.

\subsection{Ablation study: CFG}
\label{subsubsec:cfg_vs_no_cfg}

\begin{figure}[H]
\centering
\includegraphics[width=0.83\textwidth]{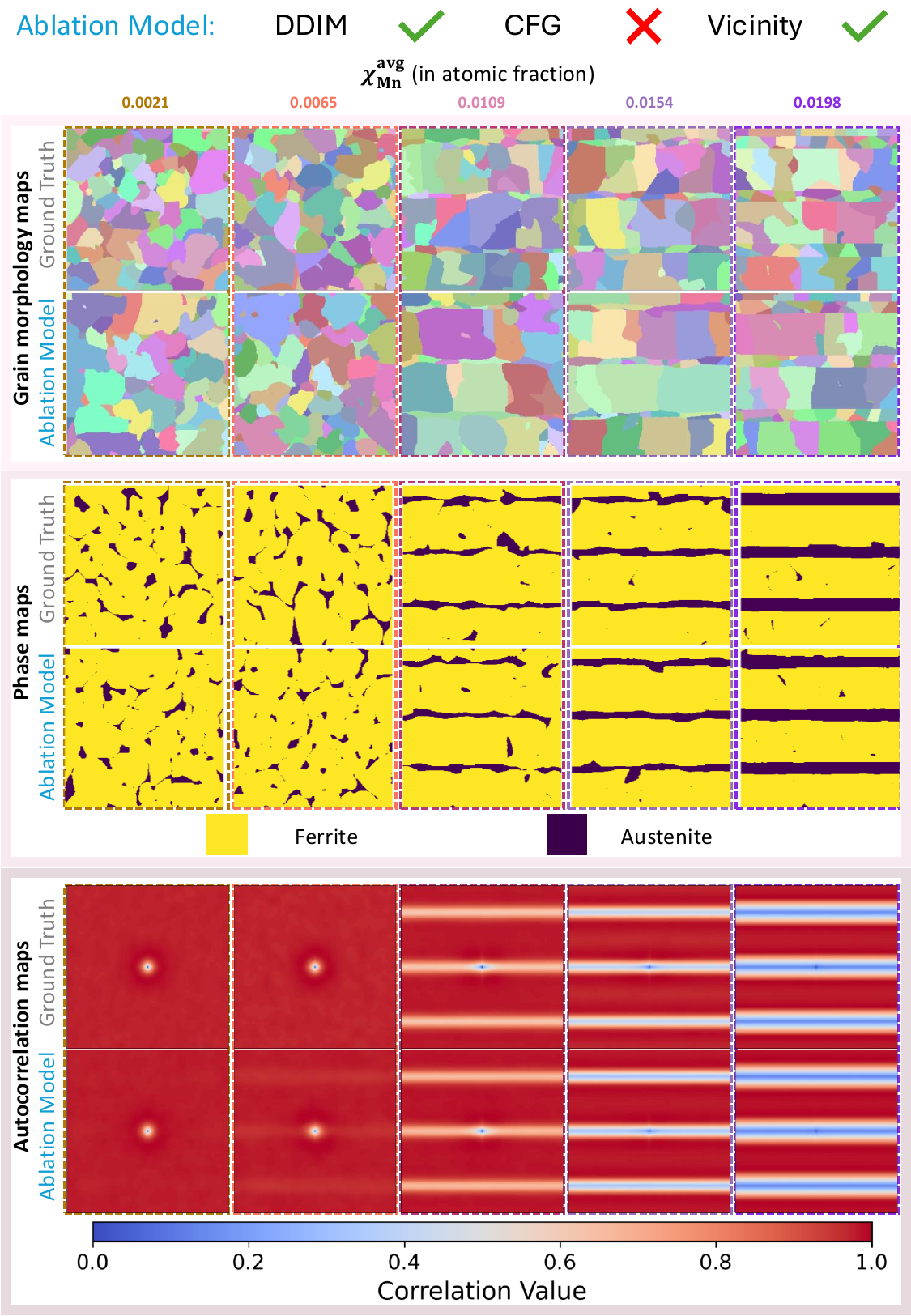}
\caption{\textbf{CFG ablation:} Qualitative comparison of microstructures between ground truth and CFG-ablation model at specific $\chi^\text{avg}_\text{Mn}$ values (indicated above the top panel) not present in the training dataset. \textit{Top panel:} grain morphology maps. \textit{Middle panel:} phase maps. \textit{Bottom panel:} autocorrelation maps.  The CFG-ablation model shows inaccuracies in the autocorrelation maps indicating issues with the banding formation.}
\label{fig:combined_qual_no_CFG}
\end{figure}

\begin{figure}[H]
\centering
\includegraphics[width=0.83\textwidth]{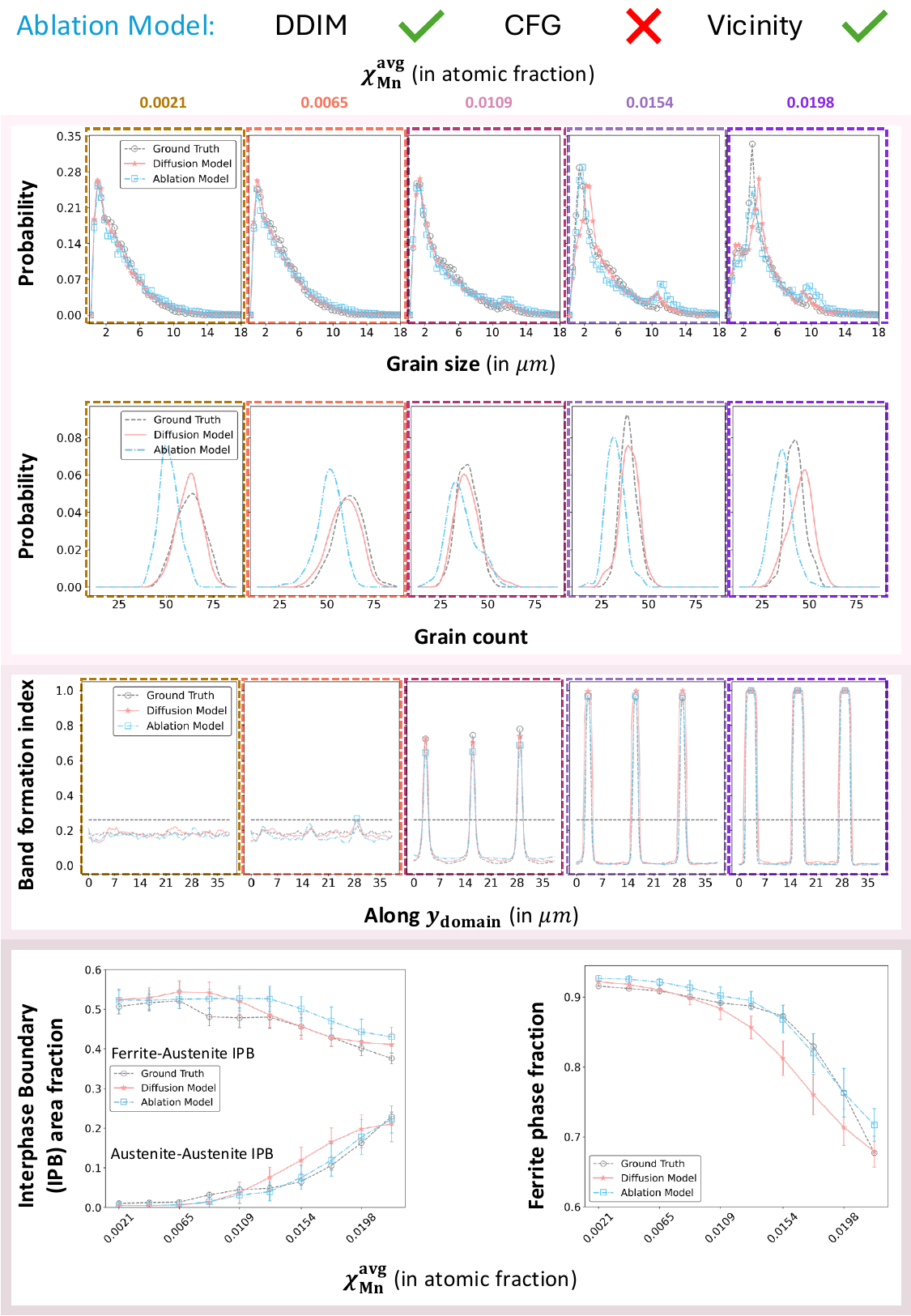}
\caption{\textbf{CFG ablation:} Quantitative comparison of microstructure statistics between ground truth (gray dashed line with circles) and CFG-ablation model (red line with filled stars) at specific $\chi^\text{avg}_\text{Mn}$ not present in the training dataset. The $\chi^\text{avg}_\text{Mn}$ values for the top and middle panels are indicated above the top panel, whereas in the bottom panel they are shown along the horizontal axes of the respective plots. \textit{Top panel:} ferrite grain size and ferrite grain count distributions.  \textit{Middle panel:}  band formation index along $y_{\text{domain}}$, i.e., the y-direction inside the microstructure domain. \textit{Bottom panel:} ferrite-austenite and austenite-austenite IPB area fractions and ferrite phase fraction vs.~$\chi^\text{avg}_\text{Mn}$. Key differences are observed in band formation index and ferrite grain count distribution.}
\label{fig:combined_quant_no_CFG}
\end{figure}

We disable CFG by setting the dropout probability to an extremely small value, $p_{\text{drop}} = 10^{-6} \ll 1$, effectively eliminating the activation of the unconditional model during training. Similarly, during the sampling stage, we set $\Gamma=1$  to suppress the contribution from unconditional generation in \eqref{eq:CFG}. For reference, the diffusion model outside the CFG ablation study uses $p_{\text{drop}} = 0.1$ and $\Gamma = 1.5$. We present qualitative and quantitative comparisons of the diffusion model with CFG and without CFG (referred to as the ablation model) in Figure \ref{fig:combined_qual_no_CFG} and Figure \ref{fig:combined_quant_no_CFG}, respectively.

The generated grain morphology and phase maps appear accurate, but the autocorrelation maps reveal inaccuracies in manganese band formation at $\chi^{\text{avg}}_{\text{Mn}} = 0.0065$. This discrepancy is further supported by comparing the band formation index, where the ablation model predicts an earlier onset of manganese banding for the given $\chi^{\text{avg}}_{\text{Mn}}$ values. Additionally, the ablation model underestimates the overall ferrite grain counts, as indicated by the leftward shift in the corresponding KDEs relative to the ground truth and the diffusion model with CFG. On the other hand, the ablation model shows a better match with the ground truth in austenite-austenite IPB area fraction and ferrite phase fraction. We attribute this to the underestimated ferrite grain counts and overestimated ferrite grain sizes compensating for the mismatch in IPB area and ferrite phase fractions of the diffusion model and thereby yielding apparently good agreement in those quantities. These inconsistent results for the ablation model are not indicative of a well-trained performance. Consequently, we favor CFG for promoting conditional generation and the overall quality of the generated microstructures.

\begin{figure}[H]
\centering
\includegraphics[width=0.83\textwidth]{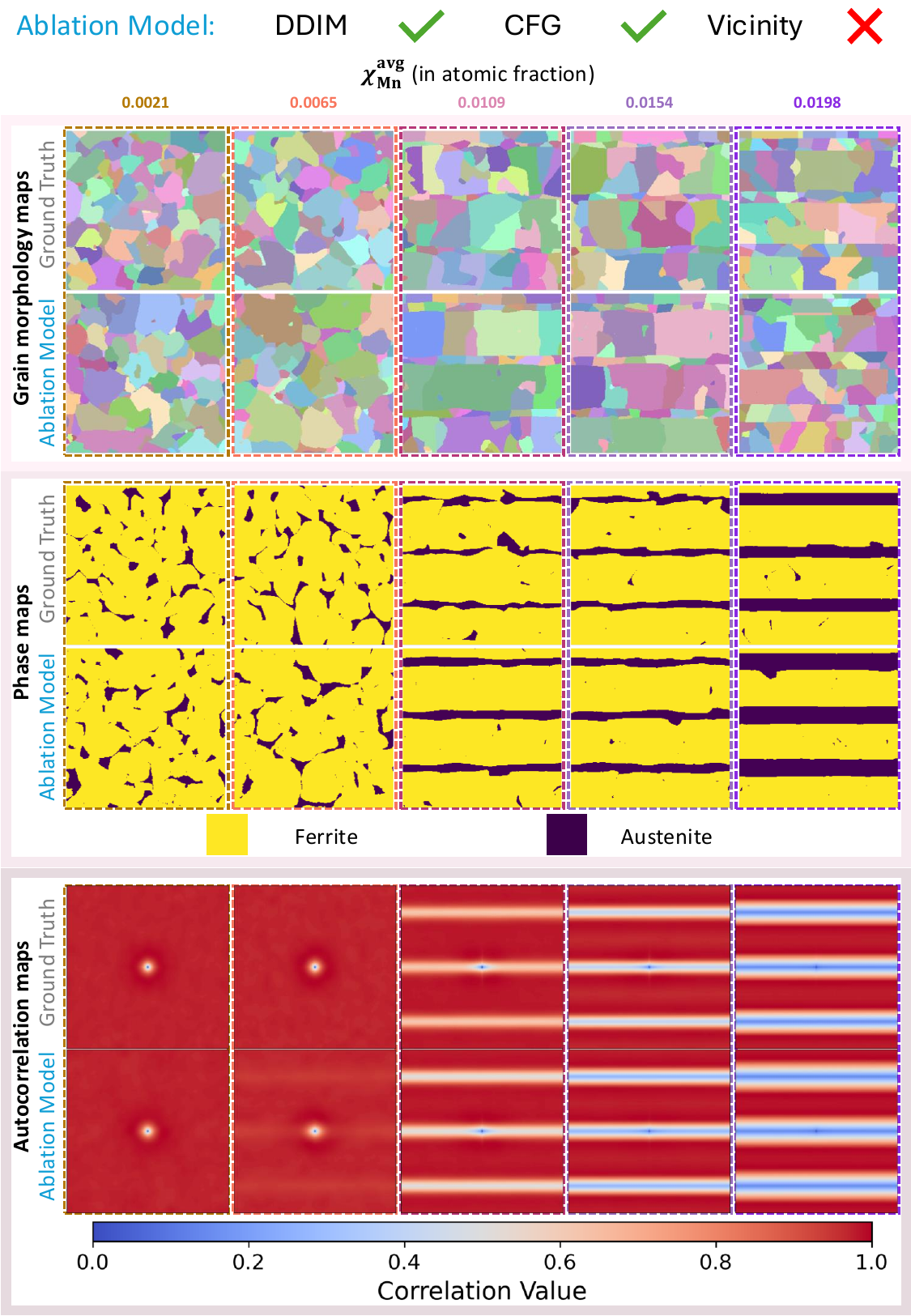}
\caption{\textbf{Vicinal loss ablation:} Qualitative comparison of microstructures between ground truth and vicinal-loss-ablation model at specific $\chi^\text{avg}_\text{Mn}$ values (indicated above the top panel) not present in the training dataset. \textit{Top panel:} grain morphology maps. \textit{Middle panel:} phase maps. \textit{Bottom panel:} autocorrelation maps.  The vicinal-loss-ablation model shows inaccuracies in the autocorrelation maps indicating issues with the banding formation.}
\label{fig:combined_qual_no_vicinal}
\end{figure}

\begin{figure}[H]
\centering
\includegraphics[width=0.83\textwidth]{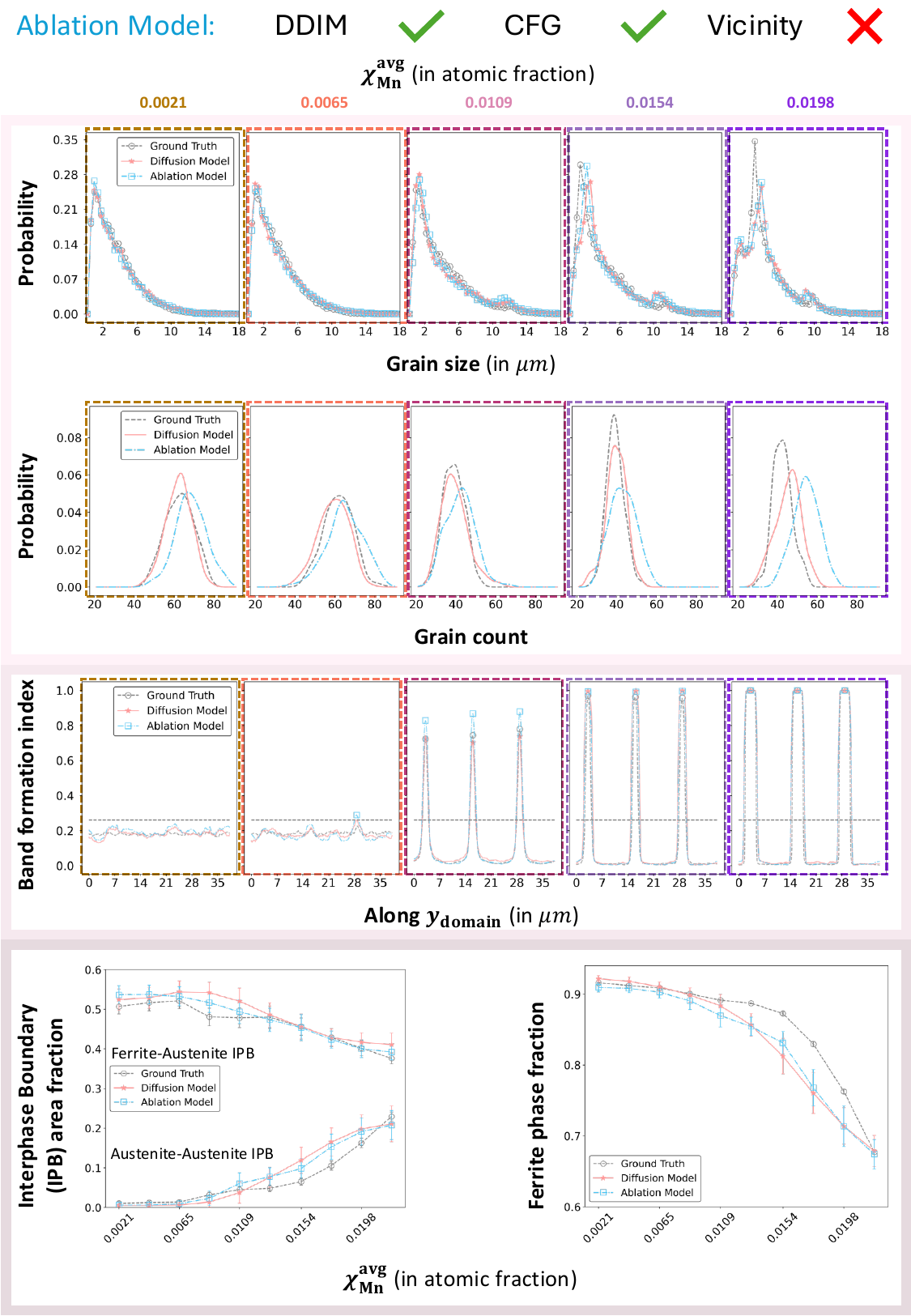}
\caption{\textbf{Vicinal loss ablation:} Quantitative comparison of microstructure statistics between ground truth (gray dashed line with circles) and vicinal-loss-ablation model (red line with filled stars) at specific $\chi^\text{avg}_\text{Mn}$ not present in the training dataset. The $\chi^\text{avg}_\text{Mn}$ values for the top and middle panels are indicated above the top panel, whereas in the bottom panel they are shown along the horizontal axes of the respective plots. \textit{Top panel:} ferrite grain size and ferrite grain count distributions.  \textit{Middle panel:}  band formation index along $y_{\text{domain}}$, i.e., the y-direction inside the microstructure domain. \textit{Bottom panel:} ferrite-austenite and austenite-austenite IPB area fractions and ferrite phase fraction vs.~$\chi^\text{avg}_\text{Mn}$. Key differences are observed in band formation index and ferrite grain count distribution.}
\label{fig:combined_quant_no_vicinal}
\end{figure}

\begin{figure}[H]
\centering
\includegraphics[width=0.83\textwidth]{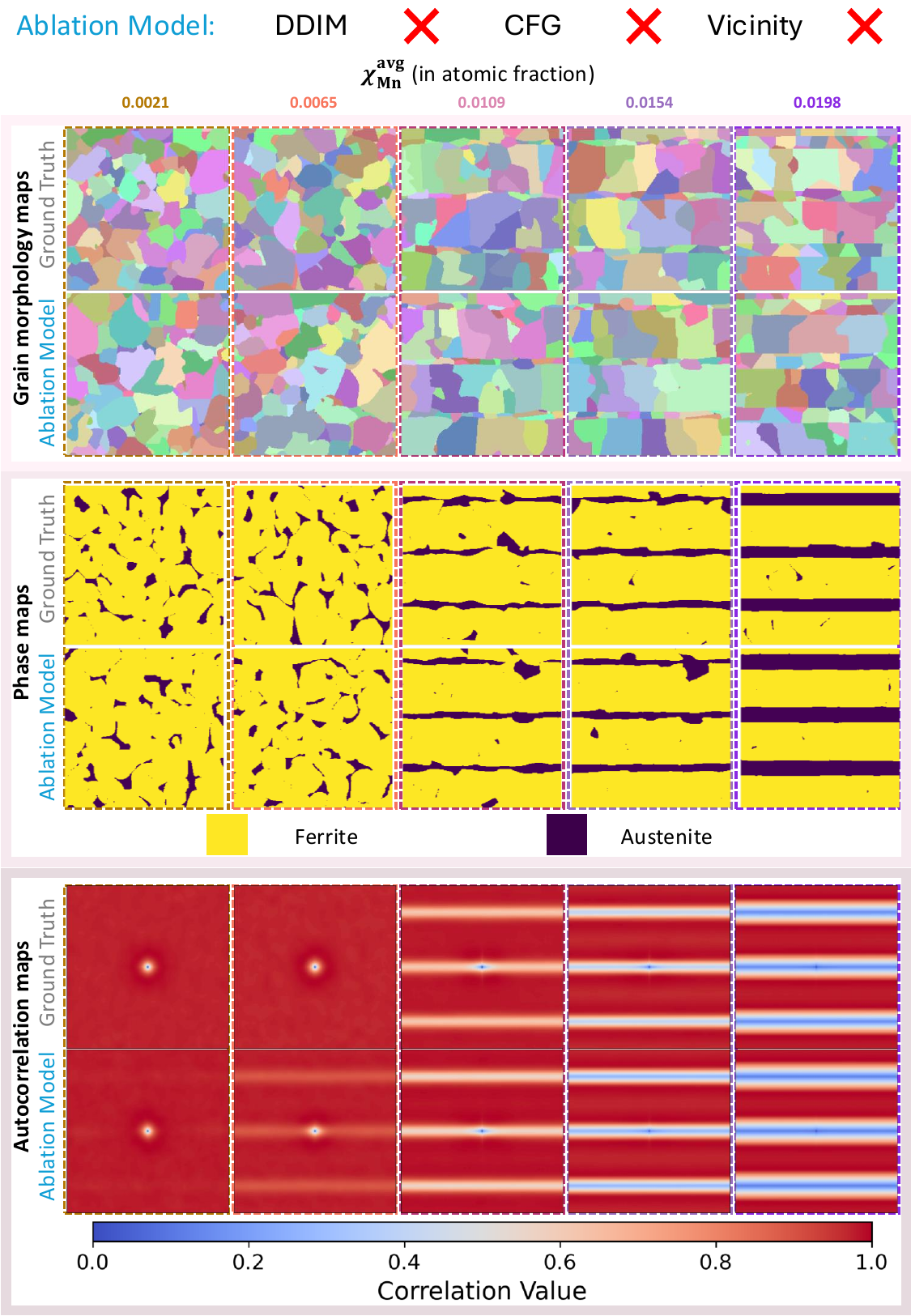}
\caption{\textbf{DDIM, CFG, and vicinal loss ablation:} Qualitative comparison of microstructures between ground truth and ablation model at specific $\chi^\text{avg}_\text{Mn}$ values (indicated above the top panel) not present in the training dataset. \textit{Top panel:} grain morphology maps. \textit{Middle panel:} phase maps. \textit{Bottom panel:} autocorrelation maps.  The ablation model shows significant inaccuracies in the autocorrelation maps indicating issues with the banding formation, particularly at $\chi^{\text{avg}}_{\text{Mn}} = {0.0065}$ and $0.0154$.}
\label{fig:combined_qual_all_off}
\end{figure}

\begin{figure}[H]
\centering
\includegraphics[width=0.83\textwidth]{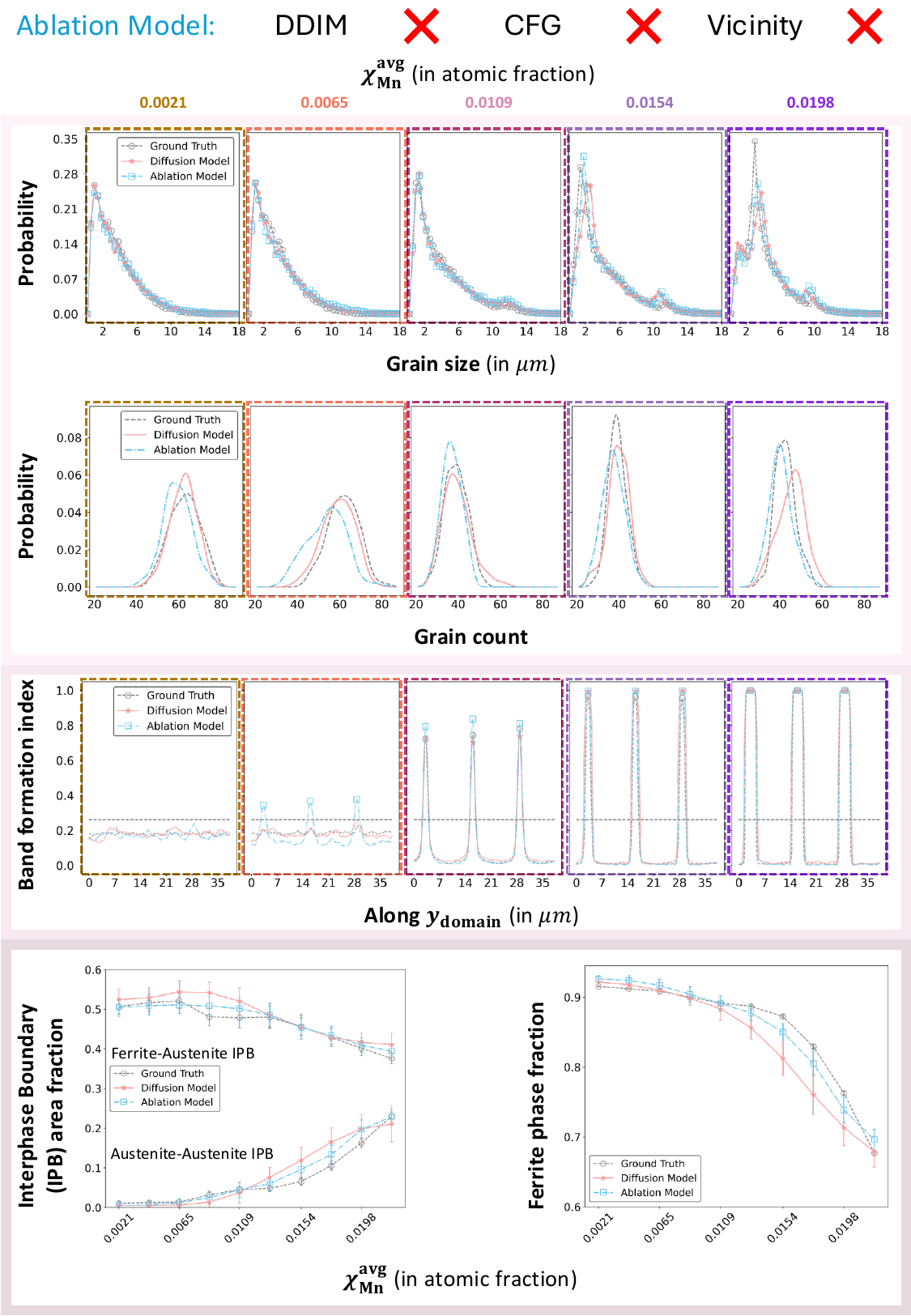}
\caption{\textbf{DDIM, CFG, and vicinal loss ablation:} Quantitative comparison of microstructure statistics between ground truth (gray dashed line with circles) and ablation model (red line with filled stars) at specific $\chi^\text{avg}_\text{Mn}$ not present in the training dataset. The $\chi^\text{avg}_\text{Mn}$ values for the top and middle panels are indicated above the top panel, whereas in the bottom panel they are shown along the horizontal axes of the respective plots. \textit{Top panel:} ferrite grain size and ferrite grain count distributions.  \textit{Middle panel:}  band formation index along $y_{\text{domain}}$, i.e., the y-direction inside the microstructure domain. \textit{Bottom panel:} ferrite-austenite and austenite-austenite IPB area fractions and ferrite phase fraction vs.~$\chi^\text{avg}_\text{Mn}$. Key differences are observed in band formation index and ferrite grain count distribution.}
\label{fig:combined_quant_all_off}
\end{figure}

\subsection{Ablation study: vicinal loss}
\label{subsubsec:vicinity_vs_no_vicinity}

We disable vicinity-based model training (introduced in Section \ref{subsec:vicinal-loss}) by setting the hyperparameter vicinity $\kappa=0$ to prevent the model from considering microstructure images in the vicinity of conditions not present in the dataset $\calD$. For reference, the diffusion model outside the vicinal loss ablation study uses $\kappa=0.00024$. We present qualitative and quantitative comparisons of the diffusion model with vicinal loss and without vicinal loss (referred to as the ablation model) in Figure \ref{fig:combined_qual_no_vicinal} and Figure \ref{fig:combined_quant_no_vicinal}, respectively.

Similar to the previous ablation study, ablating the vicinal loss produces accurate grain morphology and phase maps but leads to inaccuracies in manganese band formation. More specifically, inaccuracies occur at $\chi^{\text{avg}}_{\text{Mn}} = 0.0065$ and $0.0154$ in the autocorrelation maps, and at $\chi^{\text{avg}}_{\text{Mn}} = 0.0065$ and $0.0109$ in the band formation index. Additionally, the ferrite grain count KDEs remain largely unresponsive to conditioning, showing a larger mismatch in the mean and variance compared to the diffusion model KDEs, thereby underscoring the significance of the vicinal loss. In \ref{sec:dataset_ablation}, we perform a data ablation study in which the amount of training data is progressively reduced, thereby increasing data sparsity and emphasizing the role of the vicinal loss under such conditions.

\subsection{Ablation study: DDIM, CFG, and vicinal loss}
\label{subsubsec:all_on_vs_all_off}

We consider an ablation model in which all previously introduced strategies are disabled, namely DDIM, CFG, and the vicinal loss. We then perform qualitative and quantitative comparisons with the diffusion model in Figure \ref{fig:combined_qual_all_off} and Figure \ref{fig:combined_quant_all_off}, respectively.

We observe larger differences in the autocorrelation maps at $\chi^{\text{avg}}_{\text{Mn}} = 0.0065$ and $0.0154$ than in  the previous ablation models. Similarly, issues in the band formation index at $\chi^{\text{avg}}_{\text{Mn}} = 0.0065$ and $0.0109$ represent an early onset of banding which makes ablation model more prone to inaccurate results. The ablation model also exhibits inaccurate trends in ferrite grain count KDEs making the training and sampling configuration is inadequate in learning the stochasticity within microstructure images. These results indicate that the proposed methods contribute meaningfully to enabling the continuous conditional diffusion model to generate high-fidelity microstructures while maintaining efficient training and sampling.

\begin{figure}[H]
\centering
\includegraphics[width=0.8\textwidth]{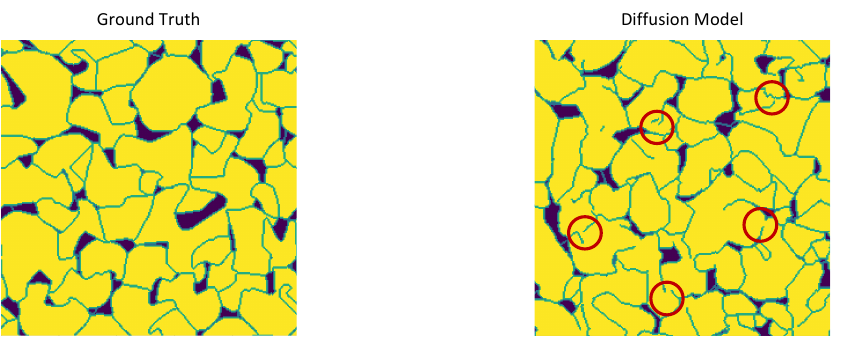}
\caption{Representative example of discrepancies (highlighted by red circles) associated with open grain boundaries in microstructures generated by the diffusion model, with a ground-truth microstructure provided for reference.}
\label{fig:limitations}
\end{figure}

\subsection{Limitations: open grain boundaries}
\label{subsec:limitations}

Although the presented continuous conditional diffusion models show strong qualitative and quantitative agreement with the ground truth microstructures, we identify a key limitation related to \textbf{open grain boundaries} in the generated microstructures. The predicted grain boundaries are sometimes disconnected or inadequately formed, leading to incorrect grain detection (see Figure \ref{fig:limitations}). This, in turn, skews the grain size and count distributions and results in inaccurate estimation of IPB area fractions and phase fractions. To the best of our knowledge, this issue has not been addressed in existing works, which often do not explicitly analyze grain boundary descriptors such as IPB area fraction, and therefore remains an open problem. Potential improvements may involve incorporating additional inductive biases or modifying the model architecture to better capture grain boundary information, for example by introducing physics-informed regularization \cite{BastekSunKochmann2024} into the diffusion framework.

\section{Conclusion}
\label{sec:Conclusion}

We presented a modeling framework based on continuous conditional denoising diffusion for generating microstructures (represented as images) as a function of process conditions. The diffusion model comprises a forward noising process and a trainable reverse denoising process guided by process conditions. Building upon classical denoising diffusion models for generative tasks, we introduced techniques to improve data and computational efficiency. Specifically, we proposed the use of vicinal loss to improve data efficiency during training when sampling from continuous process conditions, classifier-free guidance to improve the quality of the generated microstructures, and a denoising diffusion implicit model to increase the sampling speed for microstructure generation.

The diffusion model generated high-fidelity microstructure images, accurately capturing grain morphology, phase distribution, and spatial autocorrelation. Statistical metrics such as ferrite grain size distribution, band formation index, interphase boundary area fraction, phase fraction, and phase-specific grain counts showed strong agreement with the ground truth. The results validated the capability of the diffusion model to capture physical phenomena governed by process conditions, such as the effect of average manganese concentration on austenite banding in low-carbon steel microstructures. Ablation studies demonstrated that vicinal loss, classifier-free guidance, and their combination all lead to considerable improvements in the generated microstructures, as indicated by both qualitative and quantitative analyses relative to the ground truth. In parallel, the denoising diffusion implicit model enabled significant improvements in runtime for microstructure generation.

While we benchmarked these methods on two-dimensional low-carbon steel microstructure images conditioned on average manganese concentration, the proposed framework is sufficiently general to extend to other microstructure systems and process conditions. Possible extensions include data-efficient three-dimensional microstructure generation, coupling with property simulators such as those based on crystal plasticity of representative volume elements of generated microstructures to establish process-structure-property relationships, and fine-tuning on experimental data, which will be the subject of future work.

\section{Declaration of competing interest}
\label{sec:Competing Interests}
The authors declare that they have no known competing financial interests or personal relationships that could have appeared to influence the work reported in this paper.

\section{Acknowledgment}
\label{sec:Acknowledgment}
This research is part of the DEPMAT project (with project number P20-22 / N21022) of the research programme Perspectief which is partly financed by the Dutch Research Council (NWO). It is also part of the Partnership Program of the Materials innovation institute M2i (\hyperlink{www.m2i.nl}{www.m2i.nl}). We also thank Jan-Hendrik Bastek for insightful discussions during the development of the diffusion modeling framework.

\section{Data availability}
\label{sec:Data availability}
{ \textit{The datasets will be made available at the time of publication.}}

\section{Code availability}
\label{sec:Code availability}
{ \textit{The code used in this work will be made available at the time of publication.}}

\bibliographystyle{unsrtnat}
\bibliography{Bib}
\newpage
\appendix

\section{Dataset generation}
\label{sec:Xmn_distribution}

\begin{figure}[hbt!]
\centering
\includegraphics[width=0.65\textwidth]{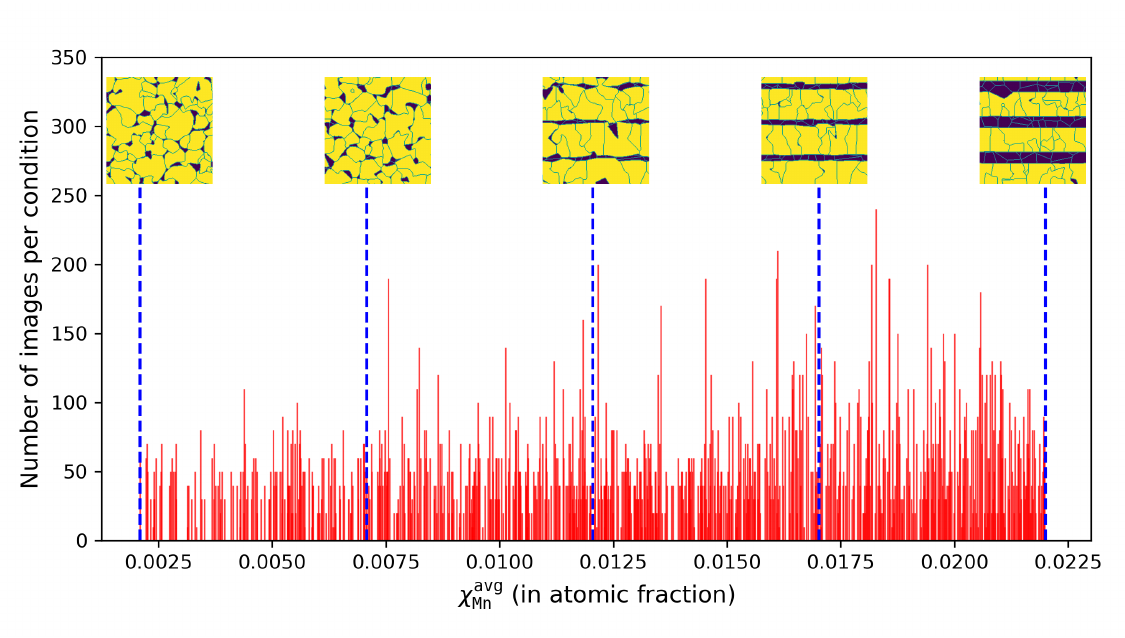}
\caption{Distribution of number of images per condition in the training dataset.}
\label{fig:dataset_distribution}
\end{figure}

To minimize the high computational cost of dataset generation, we employ a biased sampling strategy for the process condition, i.e., the average manganese concentration, $\chi_{\text{Mn}}^{\text{avg}}$, which lies in the range of $0.0021$ to $0.0220$ atomic fraction. We observe that the manganese banding effect in the microstructures is triggered at high $\chi_{\text{Mn}}^{\text{avg}}$, which is an important microstructural feature of interest. Therefore, we first partition the range of $\chi_{\text{Mn}}^{\text{avg}}$ into two intervals: a low region $[0.0021, 0.0100]$ and a high region $[0.0100, 0.0220]$. For samples drawn from the low region, we apply a rejection criterion with probability $p_{\text{reject}} = 0.5$. In other words, we retain a sample only if a randomly generated number $u \sim \calU(0, 1)$ satisfies $u \geq p_{\text{reject}}$, else we exclude it. No rejection criterion is applied when the sample belongs in the high region. The bias favors sampling of larger values more frequently in order to trigger the manganese banding effect in the microstructures. By designing the sampling of process conditions in this manner, we observe sufficient diversity in the morphologies and properties of the microstructure images (see Figure \ref{fig:dataset_distribution} for representative examples), while keeping the overall computational cost of dataset generation reasonable.

Through the above strategy, we generate a set of $1{,}000$ conditions. For each condition, the thermodynamic tables required for cellular automata simulations are generated using ThermoCalc \citep{ThermoCalc2025}. From this set, we then randomly sample $4{,}000$ times and generate $10$ microstructures for each sample using CASIPT-based cellular automata \citep{Bos2010,Bos2011}, each with a different random seed. This results in a dataset of $N = 40{,}000$ condition–microstructure pairs. We perform a $80\%-20\%$ split of the dataset into training and validation sets, see \ref{sec:loss_curves} for the loss curves during training of the diffusion model.

We preprocess the images as per the representation definition detailed in Section \ref{subsec:problem_setting}. Since the microstructures are generated at a high resolution of $256 \times 256$ pixels (necessary to capture grain boundaries), the required memory becomes substantial. To reduce the memory footprint, the dataset is stored in a compressed \texttt{float16} format. Due to the reduced  precision of \texttt{float16}, distinct $\chi_{\text{Mn}}^{\text{avg}}$ values at \texttt{float64} precision become quantized to identical representable numbers. This quantization reduces the number of unique $\chi_{\text{Mn}}^{\text{avg}}$ values from the original $1{,}000$ samples down to $746$ samples. This leads to multiple distinct conditions being associated with larger number of microstructure images. We treat this data corruption due to precision loss as an additional challenge for the ML model tasked with generating high-fidelity microstructure images across all process conditions sampled continuously over the range of $\chi_{\text{Mn}}^{\text{avg}}$.

\section{Noise scheduler}
\label{sec:noise_scheduler}

The noise schedule in \eqref{eq:fwd_process} is governed by $\beta_t$. We adopt the cosine scheduler \citep{Nichol2021} to define $\beta_t$ as a function of $t\in\{1,\dots,T\}$ as follows:

\begin{equation}
\beta_t = 1 - \frac{\bar{\alpha}_{t}}{\bar{\alpha}_{t-1}}, \quad \text{where} \quad \bar{\alpha}_{t} = \cos\left(\frac{t/T + s}{1 + s} \cdot \frac{\pi}{2}\right)^2
\end{equation}

\cite{Nichol2021} introduce an offset hyperparameter $s$ to avoid very small $\beta_t$ values near $t = 0$, see \ref{sec:parameters} for parameter value. Furthermore, $\beta_t$ is clipped to a maximum of $0.999$ to prevent numerical singularities at $t = T$.

\section{Training and validation loss curves}
\label{sec:loss_curves}

\begin{figure}[hbt!]
\centering

\includegraphics[width=0.8\textwidth]{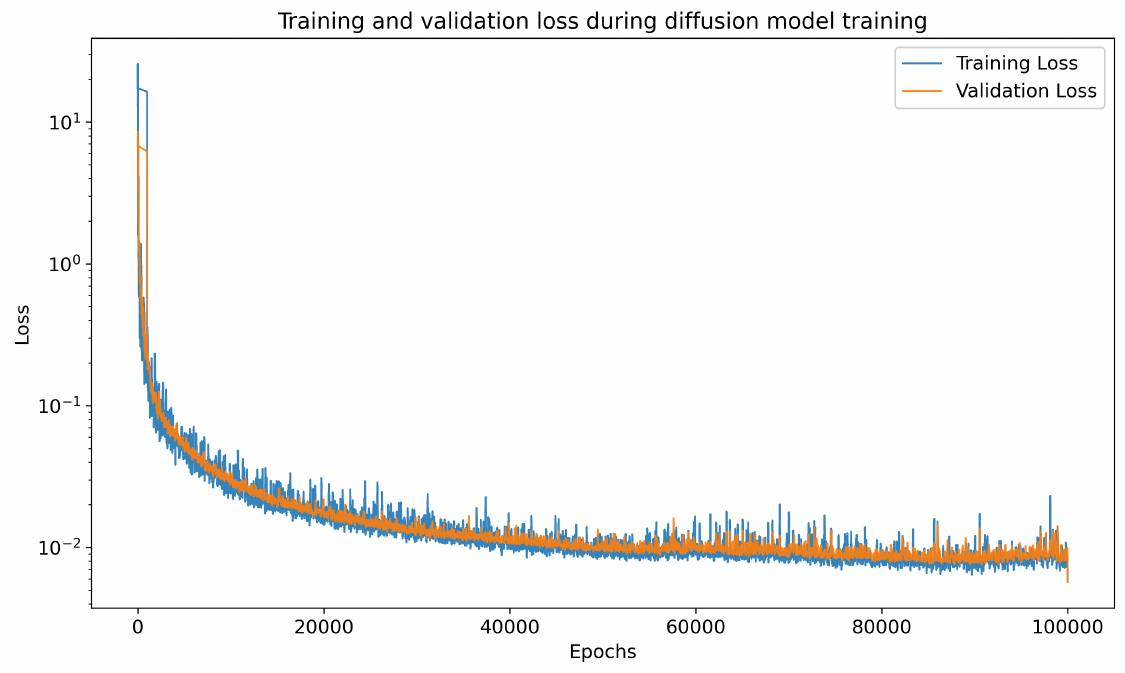}
\caption{Training and validation loss of the diffusion model (presented in Section \ref{subsubsec:qual_results_best} and Section \ref{subsec:quant_results_best}) for 100{,}000 epochs.}
\label{fig:loss_curve}
\end{figure}

Figure \ref{fig:loss_curve} shows the loss curves for the training and validation of the diffusion model discussed in Section \ref{subsubsec:qual_results_best} and Section \ref{subsec:quant_results_best}. We observe a steady decline in the loss indicating stable training.

\section{Dataset Ablation Studies}
\label{sec:dataset_ablation}

We perform dataset ablation studies to further assess the model capabilities and compare the quality of generated microstructure images with the ground truth as well as the diffusion model presented and discussed in Section \ref{subsubsec:qual_results_best} and Section \ref{subsec:quant_results_best}. Firstly, we reduce the total \textbf{number of process conditions} passed during diffusion model training to observe the effect of having fewer continuous conditions for generative modeling application. Secondly, we reduce the total \textbf{number of microstructure images} passed during diffusion model training to observe the consequence on the fidelity of the microstructure images generated. We train the ablation models after shrinking the dataset and sample microstructure images for 5 representative conditions not present in the training dataset: $\chi^\text{avg}_\text{Mn}\in\{0.0021, 0.0065, 0.0109, 0.0154, 0.0198\}$ (in atomic fraction units).

\subsection{Number of process conditions}

\begin{figure}
\centering
\includegraphics[width=0.83\textwidth]{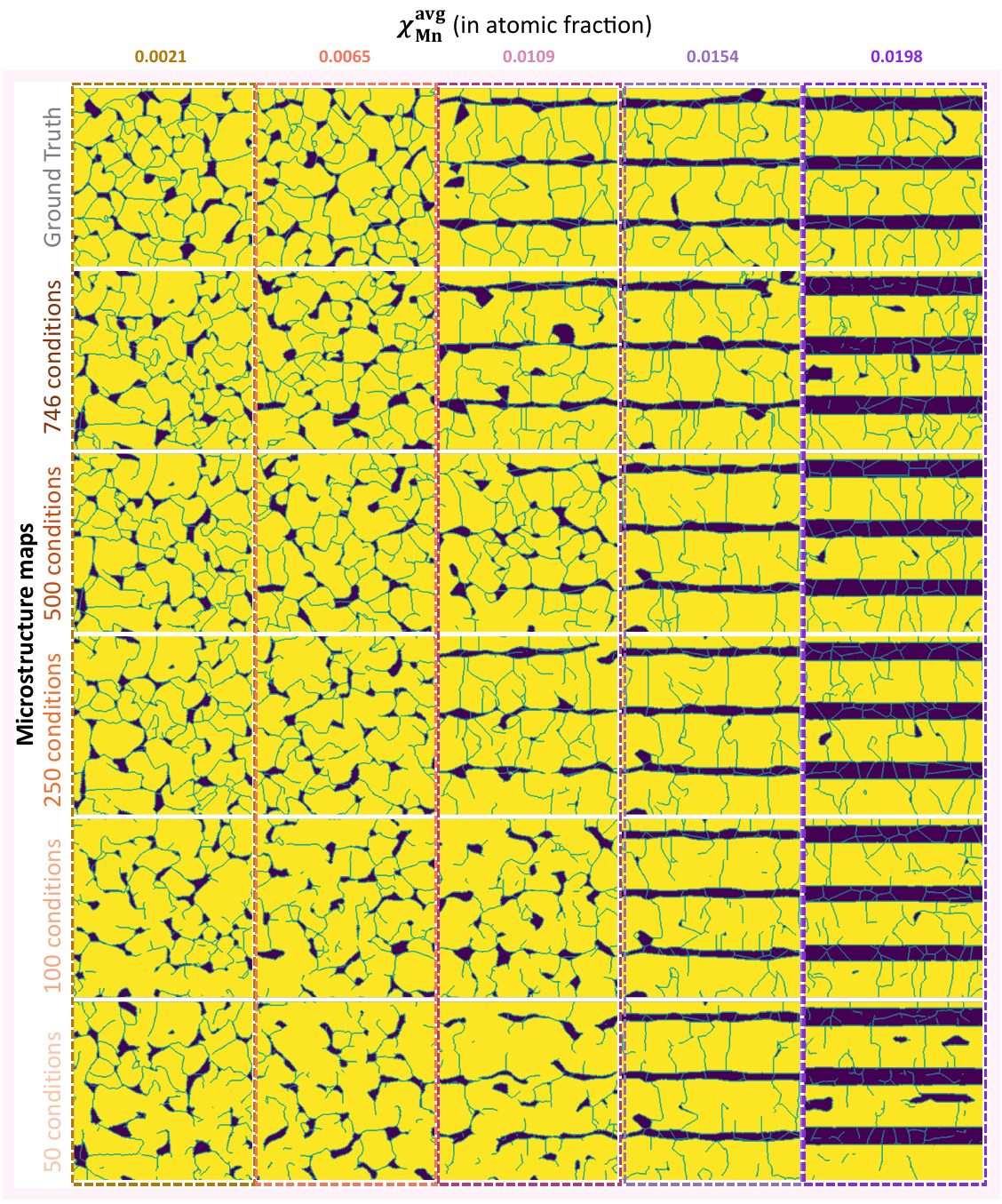}
\caption{Comparison of generated microstructure maps between the ground truth, diffusion model with 746 $\chi^\text{avg}_\text{Mn}$ conditions and ablation models with 500, 250, 100, and 50 $\chi^\text{avg}_\text{Mn}$ conditions respectively, at $\chi^\text{avg}_\text{Mn}\in\{0.0021, 0.0065, 0.0109, 0.0154, 0.0198\}$ (in atomic fraction units) not present in the training datasets.}
\label{fig:combined_n_cond}
\end{figure}

We sort all unique process conditions and then uniformly sample a subset based on the number of conditions used to train the ablation models. All the microstructure images corresponding to the selected conditions are included during the ablation model training while rest of images are excluded. The original diffusion model is trained with 746 conditions whereas we pick four ablation models trained on 500, 250, 100, and 50 conditions respectively, see Figure \ref{fig:combined_n_cond} for the microstructure images.

We observe progressively increasing degradation in the generated microstructure maps as the number of process conditions in the dataset decreases. While the diffusion model trained on all 746 conditions closely reproduces the ground truth microstructures, the ablation models introduce noticeable artifacts, including inconsistencies in grain boundaries and phase interfaces. At 500 conditions, the grains retain partial continuity but exhibit minor boundary irregularities but exhibit poor manganese banding (particularly at $\chi^\text{avg}_\text{Mn} = 0.0109$). With 250 conditions, more pronounced open boundaries appear, and at 100 and 50 conditions, both interphase grain boundary formation and manganese banding are poorly captured. These results indicate that a sufficiently large and diverse set of process conditions is essential for accurately learning both manganese band formation and interphase grain boundary generation.

\subsection{Number of microstructure images}

\begin{figure}
\centering
\includegraphics[width=0.83\textwidth]{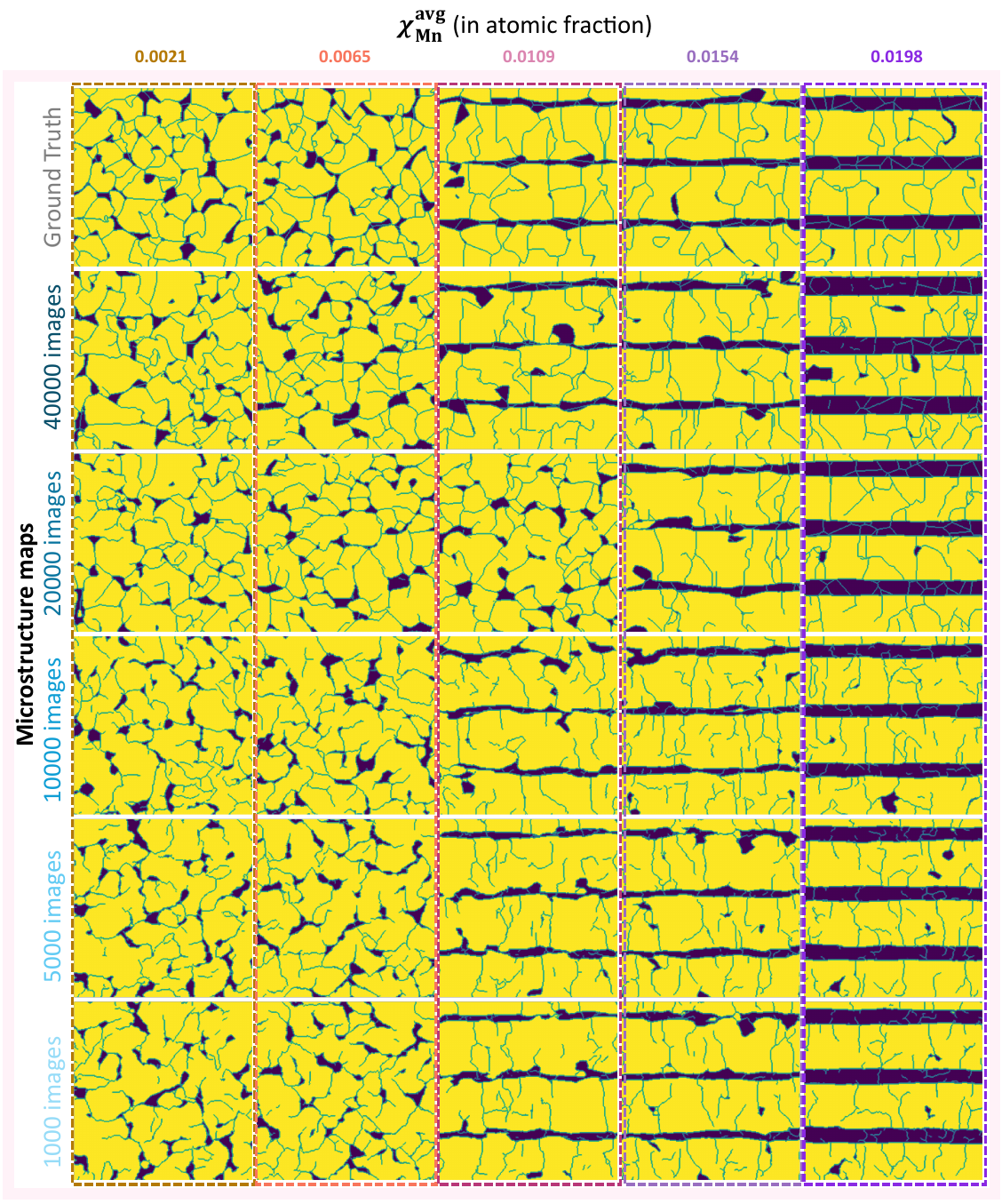}
\caption{Comparison of generated microstructure maps between the ground truth, diffusion model with $40{,}000$ microstructure images and ablation models with $20{,}000$, $10{,}000$, $5{,}000$, and $1{,}000$ microstructure images respectively, at $\chi^\text{avg}_\text{Mn}\in\{0.0021, 0.0065, 0.0109, 0.0154, 0.0198\}$ (in atomic fraction units) not present in the training datasets.}
\label{fig:combined_n_images}
\end{figure}

We control the training dataset size by setting a maximum number of images per condition. We fine-tune this parameter until the overall dataset closely includes the desired number of microstructure images. The original diffusion model is trained with $40{,}000$ images whereas we pick four ablation models trained on approximately $20{,}000$, $10{,}000$, $5{,}000$, and $1{,}000$ images respectively, see Figure \ref{fig:combined_n_images} for the representative examples of the generated microstructures. While the manganese banding is relatively unaffected, we notice a severe degradation in the interphase grain boundary generation, especially the austenite-austenite interfaces at high $\chi^\text{avg}_\text{Mn}$ values. The amount of open and incomplete grain boundaries increases as the number of microstructure images in the dataset decreases.
These results indicate that a sufficiently large number of high quality microstructure images (per condition) is needed to learn the continuous conditioning of the diffusion model.

\section{Configuration}
\label{sec:parameters}

We present all the parameters and hyperparameters used in this work along with their corresponding values in Table \ref{tab:parameter_table}.

\begin{table}[htbp]
\centering
\caption{Parameter and hyperparameter values used for data generation, model training, and sampling.}
\label{tab:parameter_table}
\begin{tabular}{|c|c|}
\hline
\textbf{Description} & \textbf{Value} \\
\hline
Number of image channels: $C$ & $1$ \\
Number of pixels: $H\times W$ & $256\times 256$ \\
Number of datapoints: $N$ & $40{,}000$ \\
Number of conditions: $P$ & $1$ \\
Average carbon concentration: $\chi^\text{avg}_\text{C}$ & $3.25 \times 10^{-3}$ atomic fraction\\
Average silicon concentration: $\chi^\text{avg}_\text{Si}$ & $1.36 \times 10^{-3}$ atomic fraction\\
Initial temperature: $T_\text{initial}$ & $940$ \textdegree C \\ 
First cooling path: $(\phi_1, \Delta t_1)$ & ($-10$ \textdegree C/s, $15$ s)\\
Second cooling path: $(\phi_2, \Delta t_2)$ & ($-12$ \textdegree C/s, $12$ s)\\
Isothermal holding time: $\Delta t_3$ & $100$ s\\
Average manganese concentration: $\chi^\text{avg}_\text{Mn}$ & $0.0021$--$0.0220$ atomic fraction\\
Amplitude of sinusoidal variation in $\chi_\text{Mn}$: $A$ & $0.25 \cdot \chi^\text{avg}_\text{Mn}$ \\
Number of sinusoidal periods in $\chi_\text{Mn}$: $N_{\text{periods}}$ & $3$ \\
Length (in pixels) of the system along Y-direction: $n_y$ & $256$ \\
Cell size per pixel: $\Delta d$ & $0.15\ \mu\text{m}$ \\
Optimizer & Adam \citep{Kingma2014} \\
Learning rate & $1 \times 10^{-5}$ \\
Epochs & $100{,}000$ \\
Training dataset size & $32{,}000$ \\
Validation dataset size & $8{,}000$ \\
Noise schedule offset: $s$ & $0.008$ \\
Number of noising steps: $T$ & $1{,}000$ \\
Number of sampling steps: $T'$ & $250$ \\
Vicinity: $\kappa$ & $0.00024$ \\
Noise standard deviation: $\sigma_{\delta}$ & $0.00069$ \\
Dimension of covariance condition embedding: $d^{\bm{\Xi}}$ & $65,536$ \\
Dimension of continuous condition embedding: $d$ & $128$ \\
Dimension of denoising step embedding: $d^t$ & $64$ \\
Dropout probability: $p_{\text{drop}}$ & $0.1$ \\
Conditional scaling: $\Gamma$ & $1.5$ \\
\hline
\end{tabular}
\end{table}

We also provide a detailed overview of the hardware specifications and runtimes for dataset generation, diffusion model training, and sampling in Table \ref{tab:specs}.

\begin{table}[htbp]
\centering
\caption{Hardware resources, software, and runtime used for different tasks.}
\label{tab:specs}
\begin{tabular}{p{4cm} p{4cm} p{4cm} p{3cm}}
\hline
\textbf{Task} & \textbf{Software} & \textbf{Hardware} & \textbf{Runtime} \\
\hline
Dataset generation & CASIPT & CPU$^1$ 4 cores  & 10--20 hours \\
Diffusion model: training & PyTorch (Python) & CPU$^1$ + GPU$^2$ & 40 hours \\
Diffusion model: sampling$^3$ & PyTorch (Python) & CPU$^1$ + GPU$^2$ & 9-10 seconds \\
\hline
\end{tabular}
\vspace{2mm}
\begin{minipage}{\textwidth}
\footnotesize
$^1$  AMD Ryzen Threadripper PRO 5965WX 3.8\,GHz processor with  256\,GB of DDR3 memory\\
$^2$ NVIDIA RTX A6000 with 48\,GB GDDR6 memory using CUDA 12.2\\
$^3$ Reported runtime corresponds to sampling a single microstructural image\\
\end{minipage}
\end{table}

\end{document}